\documentclass[seceq]{ptptex}

\usepackage{graphicx}

\title{
Effects of Magnetic Fields on Proto-Neutron Star Winds%
}

\author{
Hirotaka Ito,$^1$ Shoichi Yamada,$^1$ Kohsuke Sumiyoshi,$^2$ and Shigehiro Nagataki$^3$ %
}

\inst{
$^1$Department of Physics, Science, and Engineering, Waseda University,
Okubo, 3-4-1, Shinjuku, Tokyo 169-8555, Japan \\
 \&\\
Advanced Research Institute for Science and Engineering, Waseda
University, Okubo, Shinjuku, Tokyo 169-8555, Japan \\
$^2$Numazu College of Technology, Ooka 3600, Numazu, Shizuoka 410-8501,
Japan \\ 
  \&\\
Division of Theoretical Astronomy, National Astronomical Observatory of
Japan, 2-21-1 Osawa, Mitaka, Tokyo 181-8588, Japan \\
$^3$Yukawa Institute for Theoretical Physics, Kyoto University,
Oiwake-cho Kitashirakawa Sakyo-ku, Kyoto 606-8502, Japan
}

\abst{
 We discuss  effects of magnetic fields on proto-neutron star winds by
 performing  numerical simulation.
 We assume that the atmosphere of proto-neutron star has a
 homogenous magnetic field (ranging from $10^{12}$G to
 $5\times10^{15}$G)  perpendicular to the radial direction
 and  examine the dependence of the three key quantities
 (dynamical time scale $\tau_{dyn}$, electron fraction $Y_e$, and entropy
 per baryon $s$) for the successful r-process on the magnetic field
 strength.
 Our results show that  even  with a magneter-class field
 strength, $\sim 10^{15}$G, the feature of the wind dynamics
 varies only little from that of non-magnetic
 winds,
 and that the condition for successful r-process is not realized.   }

\begin{document}
\maketitle

\section{Introduction}

 Because of their nuclear charges and energetical disadvantages,
 elements  heavier than
 iron cannot be produced by thermonuclear reactions. However, 
 their solar abundances are much greater than those can be obtained in the
 nuclear statistical equilibrium. 
 One of the processes that contribute to synthesis of
 this heavy element  is the r-process (rapid neutron-capture
 process). With no Coulomb barrier to overcome, the seed nuclei of
 the r-process
 capture neutrons easily even at low energies, forming 
 characteristic abundance peaks at A $\sim$ 80, 130, and 195.
 The heaviest elements, in particular, such as $^{235}$U,  $^{238}$U, and
 $^{232}$Th can  be originated only in the r-process.

 In the r-process, the typical time scale of neutron capture must be much
 shorter than that of $\beta$-decay. Thus, the neutron-to-seed ratio is a
 critical parameter for the r-process and must be significantly
 large ($>$ 100) to pile up enough neutrons to produce nuclei up to and
 beyond the third peak of the abundance.
 Identifying the sites that provide with this large amount of 
 neutrons has been an attractive issue  but is still debated.

 Recent observations of r-process elements in
 ultra-metal-poor halo stars show an excellent coincidence with the solar
 r-process abundances except possibility for the lighter elements.
 This suggests that the r-process has occurred from
 early phase of the galactic evolution. 
 Thus, identifying  the r-process site should 
 give us  information on the  galactic evolution,
 the star formation rate, and the
 initial mass function.

 The studies of the r-process elements are also important because
 some of  them can be  used as cosmological chronometers.
 For example, uranium and thorium are considered to be useful, since
  they have long radiative half lives ($^{238}$U;4.468Gyr
 $^{232}$Th;24.08Gyr). The age of the metal-poor stars, the lower limit
 of the age of universe, can be estimated by
 comparing their observed abundance ratio with their theoretical
 initial production ratio.

 Among a number of the r-process sites proposed so far, the
 neutrino-heated wind from the surface of the 
 proto-neutron star (PNS) seems to be the
 most promising. The neutrino-heated wind is an outflow which is
 considered to  take place
 at the late phase of the delayed explosion scenario.

 Since this  wind scenario was proposed, the dynamics  and the
 nucleosynthetic outcome have been studied.
 Woosley et al.~\cite{rf:1}  performed numerical simulations by
 using a spherically symmetric hydrodynamics code. They  followed
 up to $\sim$18s the
 wind evolution after the trigger of the core-collapse of a
 20$M_{\odot}$ star.
 They found an outflow with significantly high entropy per baryon
 $\gtrsim$400$k_{b}$, where $k_b$ is the Boltzmann constant,  and succeeded in 
 producing the r-process solar abundance.  However an outflow with such
 a high entropy to bring a successful r-process has not been reproduced
 in other numerical simulations\cite{rf:2,rf:3}.
 Qian \& Woosley~\cite{rf:4} has done both analytical and numerical
 examination  systematically to investigate the 
 properties of the
 wind to seek for a site of r-process.
 Their
 analytic models are based on the  approximate solutions of  
 spherically symmetric steady outflows. They compared them
 with the numerical calculations. They showed clearly that the wind dynamics
 is characterized by three quantities, that is, the dynamical time
 scale $\tau_{dyn}$, entropy per baryon $s$, and electron fraction $Y_e$.  
 Although they obtained a good agreement between analytical and numerical
 results, the obtained entropy fell short of the require value
 by a factor of 2-3.

 Cardall \& Fuller~\cite{rf:5} extended the Newtonian analysis of
 Ref.~\cite{rf:4} to the general relativistic one. They  showed
 that the increase of mass-to-radius ratio of the
 PNS due to the stronger general relativistic gravity leads to an
 increase of $s$ and decrease of 
 $\tau_{dyn}$, which is favorable for the r-process.
 Further studies of the wind dynamics in the framework of general relativity
 were done analytically by Otsuki et al.~\cite{rf:6} and 
 Wanajo et al.~\cite{rf:7}, as well as numerically
 by Sumiyoshi et al.~\cite{rf:8}. In their works,
 it was  found there is a possibility of  successful r-process nucleosynthesis
 to occur with an entropy not so high as $\sim400k_b$ but with a very short
 dynamical time scale
 $\tau_{dyn}\sim6$ms. However, to realize this condition, a very small
 mass-to-radius ratio ($GM/c^{2}R\sim 0.3$) is required,
 which can not be met by a
 model with a typical neutron star (NS) mass ($\sim1.4M_{\odot}$) and 
 radius ($\sim10$km) (see, however, Terasawa et al.~\cite{rf:17}).

 Since  there had been eventually no successful model with a
 typical neutron star ($1.4M_{\odot}$, $R\sim10$km),
 effects of magnetic field on wind dynamics were investigated by
 Nagataki et al.~\cite{rf:8-5} and
 Thompson~\cite{rf:9}. Assuming a
  monopole-like configuration,  Nagataki et al.   explored 
 effects of the magnetic field and rotation analytically, but 
 could not find a suitable condition for the
 successful r-process.
 Thompson, on the other hand,
 investigated  effects of dipole  magnetic field by assuming
 that the closed magnetic field works to trap the matter, 
 he  insisted that
 the magnetized wind is more efficiently heated than in non-magnetic models and
 claimed that a model with a surface magnetic field
 $\gtrsim6\times10^{14}$G
 would bring a successful r-process nucleosynthesis. 
 Though Thompson   claimed that he 
 found a successful r-process wind model,
 he did not solve the dynamics, and it is obvious that a
 more detailed
 treatment of the  dynamics of magnetized wind is needed to obtain a
 firm conclusion. 
 Thus, in the present study, we have
 done it by detailed numerical
 simulations. As a first step of studying magnetic effect on PNS wind,
 however, we
 have considered only a radial motion of a thin surface layer with 
 a homogeneous magnetic field perpendicular to the radial direction.

 The paper is organized as follows.
 In \S2, we  explain the numerical method used in the present
 study. 
 In \S3, we show the results of our simulations. 
 The results are analyzed in \S4. Summary is given in \S5. 

\section{Numerical method}

 We consider the subsonic outflow of magnetized matter from the PNS
 surface. Although this is a genuinely multi-dimensional problem, we
 treat only the radial motion in this paper and solve  one dimensional
 magnetohydrodynamics (MHD)
 equations as explained below. This  will approximate the flow near
 the equator 
 with  magnetic fields perpendicular to the
 radial direction.
 We have employed the  code employed by Ref.~\cite{rf:8},
 which is an implicit lagrangian code for general relativistic and
 spherically symmetric hydrodynamics (Yamada~\cite{rf:10}). 
 In the following, the space-time metric
 is approximated by 
 \begin{equation}
 ds^2=e^{2\phi(t,m)}c^2dt^2-e^{2\lambda(t,m)}\biggl(\frac{G}{c^2}\biggr)^2\displaystyle{dm^2}-r^2(t,m)\displaystyle{\left(d\theta^2+\sin^2\hspace{-0.5mm}\theta\hspace{0.5mm}d\phi^2\right)},
\label{metric}
\end{equation}  
 and we have solved the conservation equations of mass, momentum and
 energy together wit Einstein equations. See Ref.~\cite{rf:10} for detail.
 Since  our  aim is to explore
 the propagation of magnetic wind, we have modified the code to
 include the time evolution of magnetic field and its dynamical feedback.

 The temporal evolution of  the magnetic field  is derived from the equation
\begin{equation}
\mathsterling_{\mathbf{u}} \mathbf{F} = 0  .
 \label{Lie}
\end{equation} 
 Here $\mathbf{F}=F_{\mu\nu}\mathbf{d}x^\mu \otimes\mathbf{d}x^\nu$ 
 is the electromagnetic field tensor, and
 $\mathsterling_{\mathbf{u}}$ denotes the Lie derivative along the four
 velocity $u^{\alpha}$. The Greek letter stands for 0$-$3.
 The above equation is valid whenever the electric fields vanish in the
 comoving frame  (Achterberg~\cite{rf:12}), since the Lie derivative
 can be written as  
\begin{equation}
\mathsterling_{\mathbf{u}} \mathbf{F} = \mathbf{d}(\mathbf{F}\cdot\mathbf{u}) + (\mathbf{d}\mathbf{F})\cdot \mathbf{u} .
 \label{Lie2}
\end{equation}  
 The first term in equation (\ref{Lie2}) vanishes from ideal MHD
 approximation, that 
 is $\mathbf{F}\cdot\mathbf{u}=F_{\mu\nu}u^{\nu}\mathbf{d}x^{\mu}=0$.
 On the other hand, the second term in  equation (\ref{Lie2}) vanishes
 independently of the ideal  MHD approximation
 by virtue of the Maxwell equation, $\mathbf{dF}=0$, which is equivalent
 to $\mathbf{F}=\mathbf{dA}$, where $\mathbf{A}$ is the vector potential.

 Since we are using the Lagrangian coordinates,
 equation (\ref{Lie}) implies that the components of the
 electronmagnetic tensor remains constant for each fluid element, which can
 be written as
\begin{equation}
F_{\mu\nu}(t_1,m) = F_{\mu\nu}(t_0,m) .
 \label{F_const}
\end{equation} 
 Here, $t_0$ and $t_1$ are arbitrary times. 
 The components of the electromagnetic fields in the local inertial frame is
 obtained by performing the following coordinate transformation at each coordinate
 point  
\begin{equation}
 \hat{F}_{\alpha\beta}(t,m)=\frac{\partial x^{\mu}}{\partial \hat{x}^{\alpha}} 
                       \frac{\partial x^{\nu}}{\partial \hat{x}^{\beta}}
                       F_{\mu\nu}(t,m) .
 \label{transform}
\end{equation} 
 Here,  $\hat{F}_{\alpha\beta}$ represents the electromagnetic tensor in
 the local inertial frame.
 The
 transformation matrix is denoted as 
 $\frac{\partial x^{\mu}}{\partial \hat{x}^{\alpha}}$, and its
 components are
\begin{equation}
 \frac{\partial x^{\mu}}{\partial \hat{x}^{\alpha}} =
\left(
\begin{array}{cccc} 
 e^{-\phi(t,m)}c^{-1} & 0 & 0 & 0 \\
 0 & e^{-\lambda(t,m)}\Bigl(\frac{G}{c^2}\Bigr)^{-1}  & 0 & 0 \\
 0 & 0 & r^{-1}(t,m) & 0 \\
 0 & 0 & 0 & r^{-1}(t,m)sin^{-1}{\theta}
\end{array}
\right) .
\label{matrix}
\end{equation} 
 Thus, from equations (\ref{F_const})-(\ref{matrix}), the magnetic
 fields in the local inertial frame at an arbitrary time is determined as  
\begin{equation}
B_{r}(t_1,m) = \frac{r^{2}(t_0,m)}{r^{2}(t_1,m)}B_{r}(t_0,m) ,
 \label{Br}
\end{equation}     
\begin{equation}
B_{\theta}(t_1,m) = \frac{r(t_0,m)}{r(t_1,m)}e^{(\lambda(t_0,m) - \lambda(t_1,m))}B_{\theta}(t_0,m) ,
 \label{Bs}
\end{equation}      
\begin{equation}
B_{\phi}(t_1,m) = \frac{r(t_0,m)}{r(t_1,m)}e^{(\lambda(t_0,m) - \lambda(t_1,m))}B_{\phi}(t_0,m) , 
 \label{Bph}
\end{equation}      
 where $B_r$, $B_{\theta}$, $B_{\phi}$ are radial,
 polar, azimuthal components of magnetic field, respectively. 
 By taking $t_0$ as the
 initial time, the time evolution of the magnetic field is determined by
 equations (\ref{Br})-(\ref{Bph}). In the present study, we only consider
 the magnetic fields perpendicular to the radial direction 
 and set $B_{r}=B_{\phi}=0$. This corresponds, for example, to the
 uniform, dipole, or  toroidal field in equator.

 The
 general relativistic equation of motion and  conservation of energy are
 modified by including dynamical feedback of magnetic field. 
 Adding the contribution of the magnetic pressure to the original equation,
 the equation of motion is given by
\begin{equation}
e^{-\phi}\frac{\partial U}{\partial t} = - \frac{\Gamma}{h} 4 \pi r^2 \biggl(\frac{\partial P}{\partial m} + \frac{\partial (B^2/8\pi)}{\partial m} + \frac{\tau q}{4 \pi r^2}\biggr) - \frac{\tilde{m}}{r^2} - 4 \pi r(P + P_{\nu}) ,
\label{momentum con2}
\end{equation}   
 where  $U$ is the radial fluid
 velocity defined as
\begin{equation}
 U = e^{-\phi} \frac{\partial r}{\partial t},
\end{equation}
 and $\Gamma$ is the general relativistic Lorentz factor defined as
\begin{equation}
 \Gamma^2 = 1 + U^2 - \frac{2 \tilde{m}}{r} ,
\end{equation} 
 and $\tau$, $h$, $q$, $\tilde{m}$, and $P_{\nu}$ are the
 inverse of the baryon mass density, the specific enthalpy, the momentum
 transfer from  neutrinos to matter,
 the gravitational mass inside radius $r$, and the pressure of
 neutrinos, respectively.
 In equation (\ref{momentum con2}), the relativistic correction of
 magnetic pressure is ignored. 
 Since the energy equation is formulated in the original code so as
 to obtain the internal
 energy density   by
 subtracting the variations of  kinetic
 energy and gravitational energy  from  the total work done by the
 matter pressure, the magnetic contribution to the kinetic energy must be
 also subtracted (magnetic pressure does no work on internal energy).
 Thus, the equation for energy conservation  is replaced by
\begin{equation}
\begin{split}
e^{-\phi}\frac{\partial \varepsilon}{\partial t} =&- \frac{1}{\Gamma} (4 \pi r^2PU)  -    \frac{h}{\Gamma^2}e^{-\phi}\frac{\partial}{\partial t} \biggl(\frac{1}{2}U^2 \biggr) -\frac{U}{\Gamma}4\pi r^2\frac{\partial (B^2/8\pi)}{\partial m} +  \frac{h}{\Gamma^2}m e^{-\phi}\frac{\partial}{\partial t} \biggl(\frac{1}{r} \biggr) \\
& -  \frac{h U}{\Gamma^2}2 \pi e^{-\phi}\frac{\partial r^2}{\partial t}(P + P_{\nu}) -   \frac{1}{\Gamma}\tau U q +  \frac{P}{\Gamma}4 \pi r \tau F_{\nu} - \tau Q   ,       
\label{energy con2}
\end{split}
\end{equation}   
 where $\varepsilon$ is the specific internal energy,
 and $F_{\nu}$  and $Q$ are  the energy flux  and the net
 heating rate of neutrinos, respectively.
 The
 general relativistic corrections from magnetic pressure are also
 neglected  in the present study for simplicity.
 They are small enough in any case.

 Although the above modifications are sufficient in principle to
 calculate the evolutions of the magnetized wind, we have made other
 technical modifications. This is because of the numerical difficulties caused
 by adding the magnetic field terms  to the original equations.
 When equations (\ref{momentum con2}) and (\ref{energy con2}) are
 employed to calculate the internal energy,
 numerical errors are accumulated,
 leading to
 unphysical solutions (such as decreasing entropy and negative
 internal energy).
 To avoid this problem,
 we have made two additional modifications.
 First, we  solve the entropy equation with the heating and cooling
 processes of neutrinos taken into account,
\begin{equation}
e^{-\phi} T\frac{\partial s}{\partial t} = -m_{u}\tau Q .
 \label{entropy2}
\end{equation}    
 Here $m_{u}$ is the atomic mass unit. Second, instead of employing the energy
 density derived from equation (\ref{energy con2}), we use the
 entropy given by equation (\ref{entropy2}) to obtain the energy
 density from the equation of state (EOS). It is true that
 equation (\ref{entropy2}) is not valid when
 a shock wave appears, but
 this is not a concern in this paper, since we investigate only the
 subsonic winds. In fact, the shock
 does not appear in this study.

 Apart from the modifications listed above, the code is
 identical with the one used by Ref.~\cite{rf:8} \  The only
 exception is the
 boundary condition for the matter pressure, which are shown below for detail.
 We  use the relativistic EOS table which was developed
 by Shen  et al.~\cite{rf:13} \  As
 the wind expands, the density becomes lower than the minimum value of 
 the original EOS table.
 We have extended the table for lower densities ($\rho \leq 10^5$ g/cm$^3$).  
 Our treatment of neutrinos is simple. Following Ref.~\cite{rf:8},
  we assume that the
 neutrino distribution is constant in time and expressed as
\begin{equation}
f_{\nu_{i}}(r) = f_{\nu_{ir}}\delta(E_{\nu_{i}}-\langle E_{\nu_{i}} \rangle) ,
 \label{distribution}
\end{equation}     
 where  
  $f_{\nu_{ir}}$ is given by (the neutrino luminosity
 $L_{\nu_{i}}$, the radius of the neutrino sphere $R_{\nu_{i}}$, and the
 average neutrino energy $\langle E_{\nu_{i}} \rangle$) as   
\begin{equation}
f_{\nu_{ir}} = \frac{2 \pi (1-x)L_{\nu_{i}}}{\langle E_{\nu_{i}}\rangle^3 R_{\nu_{i}}^2} ,
 \label{disutribution coefficient}
\end{equation}      
\begin{equation}
x = \left(1 - \frac{R_{\nu_{i}}^2}{r^2} \right)  .
 \label{x}
\end{equation}      
 The average energies are 
 taken as follows,   
\begin{equation}
\langle E_{\nu_{e}} \rangle = 10 MeV ,
\label{nue}
\end{equation}
\begin{equation}
\langle E_{\bar{\nu}_{e}} \rangle = 20 MeV ,
\label{nuebar}
\end{equation}
\begin{equation}
\langle E_{\nu_{\mu}} \rangle = 30 MeV .
\label{numu}
\end{equation}
 In the present study, we do not distinguish $\mu$, $\tau$ neutrinos and
 their anti-neutrinos with each other, so
 the identical average energy is adopted for them.
 The luminosity is assumed to be the same for all flavors and is taken as 
\begin{equation}
L_{\nu_{i}}  = 10^{51} ergs/s .
\label{Lnu}
\end{equation}
 We assume that   the
 radius of neutrino spheres 
 $R_{\nu_{i}}$ are all the same with that of PNS, $R$.

 In the present paper, we set the radius $R=10$km and gravitational
 mass $M=1.4M_{\odot}$, which are typical values for observed NS.
 A very thin surface layer 
 (initially $\sim 400$ m) of PNS
 which contains a baryon mass of $\sim 10^{-6}M_{\odot}$  
 is constructed on the numerical grid.  Within this layer, 
 100 zones are uniformly spaced in the
 baryon mass coordinates ($\sim 10^{-8}M_{\odot}$ per zone).

 Since the wind is expected to be blown from a nearly hydrostatic PNS
 surface, the initial density profile of   matter is determined by solving
 the Oppenheimer-Volkoff equation in this surface layer by setting 
 the  density at the innermost mesh
 as $2\times10^{10}$g/cm$^3$ and  assuming a constant
 value for the temperature $T$ and the electron fraction $Y_{e}$ in
 the whole layer.
 In the present study, we take $T=3$MeV and $Y_{e}=0.25 $.  
 By setting the matter pressure at the outer boundary
 (see equation (\ref{Pbnd}))
 by hand, there is a little deviation from a perfectly static configuration.
 We  run the code for a short time without 
 heating and cooling processes of neutrinos
 until a complete hydrostatic state is obtained for the given outer
 boundary condition.
 We  adopt the resulting configuration as the initial condition. 
 The hydrostatic distributions obtained by this procedure 
 are shown in figure \ref{fig:2}.  The
 electron fraction remains constant (=0.25) because no neutrino
 reactions are included.

 We have assumed a  homogenous configuration for the initial magnetic fields.
 Thus, even after adding them to the matter distribution obtained above,
 the initial  condition  remains hydrostatic.
 As mentioned previously, the direction of the magnetic field is assumed to be 
 perpendicular to the radial direction ($B_{\theta}(m)=const$,
  $B_r,B_{\phi}=0$).
 We adopt a magnetic field strength ranging from $\sim 10^{12}$ G
 (typical value 
 for an ordinary NS) to $\sim 5 \times 10^{15}$ G (suitable for a magnetar).
 Together with the non-magnetic model (model N),
 the  models in this study are listed in Table \ref{table:1}.
 The names of these models are characterized by their initial 
 plasma beta $\beta$ $\equiv P/(B^2/8\pi)$ at the outer boundary.

\begin{table}[h]
 \begin{center}
  \begin{tabular}{cllll}
 \hline
 \hline
  Model  & \phantom{0} $B_{\theta}$(G)& \phantom{0} $\beta_{in}$  &
   \phantom{0} $\beta_{out}$ & $P_b$(dyn/cm$^2$) \\[2pt]

 \hline
         &&&&\\[-10pt]

    N    & \phantom{00}   0   &                 &                & 1.0$\times$10$^{22}$\\[3pt]

 10$^6$M &1.0$\times$10$^{12}$ & 1.6$\times$10$^6$& 1.0$\times$10$^5$  &
   1.0$\times$10$^{22}$\\[3pt]

 10$^3$M & 1.0$\times$10$^{13}$  & 1.6$\times$10$^4$& 1.0$\times$10$^3$  & 1.0$\times$10$^{22}$\\[3pt]

 10M  & 1.0$\times$10$^{14}$   &1.6$\times$10$^2$  & 1.0$\times$10   &
   1.0$\times$10$^{22}$\\[3pt]

  1M  &  3.2$\times$10$^{14}$ & 1.6$\times$10  & 1.0       
   & 9.9$\times$10$^{22}$\\[3pt]

 $\frac{\textstyle{1}}{\textstyle{10}}$M  & 1.0$\times$10$^{15}$    &
   1.6  & 1.0$\times$10$^{-1}$      & 9.0$\times$10$^{21}$\\[3pt]

  $\frac{\textstyle{1}}{\textstyle{20}}$M  & 1.4$\times$10$^{15}$     & 7.9$\times$10$^{-1}$  & 5.0$\times$10$^{-2}$    & 8.0$\times$10$^{21}$\\[3pt]

  $\frac{\textstyle{1}}{\textstyle{50}}$M  & 2.2$\times$10$^{15}$     & 3.2$\times$10$^{-1}$  & 2.0$\times$10$^{-2}$    & 6.3$\times$10$^{21}$\\[3pt]

  $\frac{\textstyle{1}}{\textstyle{100}}$M  & 3.2$\times$10$^{15}$     & 1.6$\times$10$^{-1}$  & 1.0$\times$10$^{-2}$    & 4.7$\times$10$^{21}$\\[3pt]

  $\frac{\textstyle{1}}{\textstyle{200}}$M  & 4.5$\times$10$^{15}$     & 7.9$\times$10$^{-2}$  & 5.0$\times$10$^{-3}$    & 2.6$\times$10$^{21}$\\[3pt]

 \hline  
  \end{tabular}
 \end{center}
\caption{Initial models. Column 1 gives the names of the models. Column
 2 presents the strength of the initial magnetic field. Columns 3 and
 4 display, respectively, the initial plasma betas in the innermost
  and
  outermost mesh points. In column 5, matter pressure  $P_{b}$ which
 appears in  equation (\ref{Pbnd}) (the outer boundary condition)  is
 shown. $P_b$ is set so as to give the same total boundary  pressure ($P+B^2/8\pi$).}
\label{table:1} 
\end{table}

 We have employed free inner and outer boundary conditions 
 except for   radius and velocity.
 At the inner boundary,
 the velocity is set to be zero, and radius is fixed in time.

 Magnetic fields $B_{\theta}$ are defined at the mesh interfaces in the
 numerical code. 
 Their inner boundary condition is given as 
\begin{equation}
 B_{\theta I} =  B_{\theta I+1} ,
 \label{Binbnd}
\end{equation}
 where the subscript I represents the innermost mesh interface.
 The outer boundary condition for magnetic fields is given as 
\begin{equation}
 B_{\theta N} = \begin{cases} B_{\theta ex} & \text{if $B_{\theta ex} \geq 0$ and $ B_{\theta ex} < B_{\theta N-1}$}\\
 B_{\theta N-1} & \text{if $B_{\theta ex} \geq B_{\theta N-1}$}\hspace{2.7cm},  \\
 0        & \text{if $B_{\theta ex} < 0$} 
  \end{cases}
 \label{Bbnd}
\end{equation}
 where the subscript N stands for the outermost interface, and 
 $B_{ex}$ is the linear extrapolation of magnetic field given as 
\begin{equation}
B_{\theta ex} =   \frac{B_{\theta N-2}  (r_{N-1} - r_{N}) + B_{\theta N-1}  (r_{N} - r_{N-2})}{r_{N-1} - r_{N-2}} .
\label{Bex}
\end{equation}
 The third condition in equation (\ref{Bbnd}) is imposed to avoid
 the negative magnetic field, although this situation has
 not occurred in our calculations. 
 The outer boundary condition for the matter pressure is expressed 
 in a similar way as
\begin{equation}
 P_{N} = \begin{cases} P_{ex} & \text{if $P_{ex} \geq P_b$ and  $P_{ex} < P_{N-1}$}\\
 P_{N-1} & \text{if $P_{ex}\geq P_b$ and $P_{ex} \geq P_{N-1}$ }, \\
 P_b & \text{if $P_{ex} < P_b$}
  \end{cases}
 \label{Pbnd}
\end{equation}
 where $P_{ex}$ is the linear extrapolation of matter pressure multiplied
 by a reduction factor 0.94, and is given as
\begin{equation}
P_{ex} = 0.94 \times   \frac{P_{n-1}  (r_n - r_{N}) + P_{n}  (r_{N} - r_{n-1})}{r_n - r_{n-1}} .
\label{Pex}
\end{equation}
 The subscripts n and N denote, respectively, 
 the outermost  mesh center and   interface.
 $P_b$ is 
 constant which varies among the  models  (see Table \ref{table:1}). These
 values are chosen so as to give an asymptotic temperature suitable for
 r-process ($T\sim 0.1$MeV) in the
 final phase of the wind evolution. 
 In fact, we have
 set $P_b$ to give the identical total pressure
 ($P+B^2/8\pi =10^{22}$dyn/cm$^2$)
 at $t = 50$s. 
 In order to suppress oscillations at
 the boundary, we have limited the boundary pressures to give only
 negative gradients.
 The  extrapolation has been employed instead  unlike
 the previous study of Ref.~\cite{rf:8}  so as  to avoid an
 unrealistic accelerations at the outermost mesh point by
 the strong magnetic field
 ($B > 10^{15}$), which may produce a shock wave.
 The factor 0.94 which appears in equation (\ref{Pex}) was
 employed only just it minimized the deviation from the
 hydrostatic state when making initial conditions, and   
 there are no physical backgrounds. 
 It is noted that
 the prescription has
 little influence on the overall dynamics.

\section{Numerical results}
 We have  computed all the models  in Table \ref{table:1} by the same
 procedure given in \S 2. The only difference in  each model is the
 magnetic field strength and the value of $P_{b}$,  the
 boundary value of  matter pressure.
 First, we show the numerical result of the non-magnetic wind case which was
 already studied by Ref.~\cite{rf:8}  (with a little change
 in the boundary conditions
 and numerical treatments
 of the basic equations). We have done this to check for the validity of
 our numerical method and  to use the result 
 for  comparison with the magnetic
 wind solutions.

\begin{figure}

  \begin{center}
  \centerline{\includegraphics[width= 15 cm,height=15 cm]{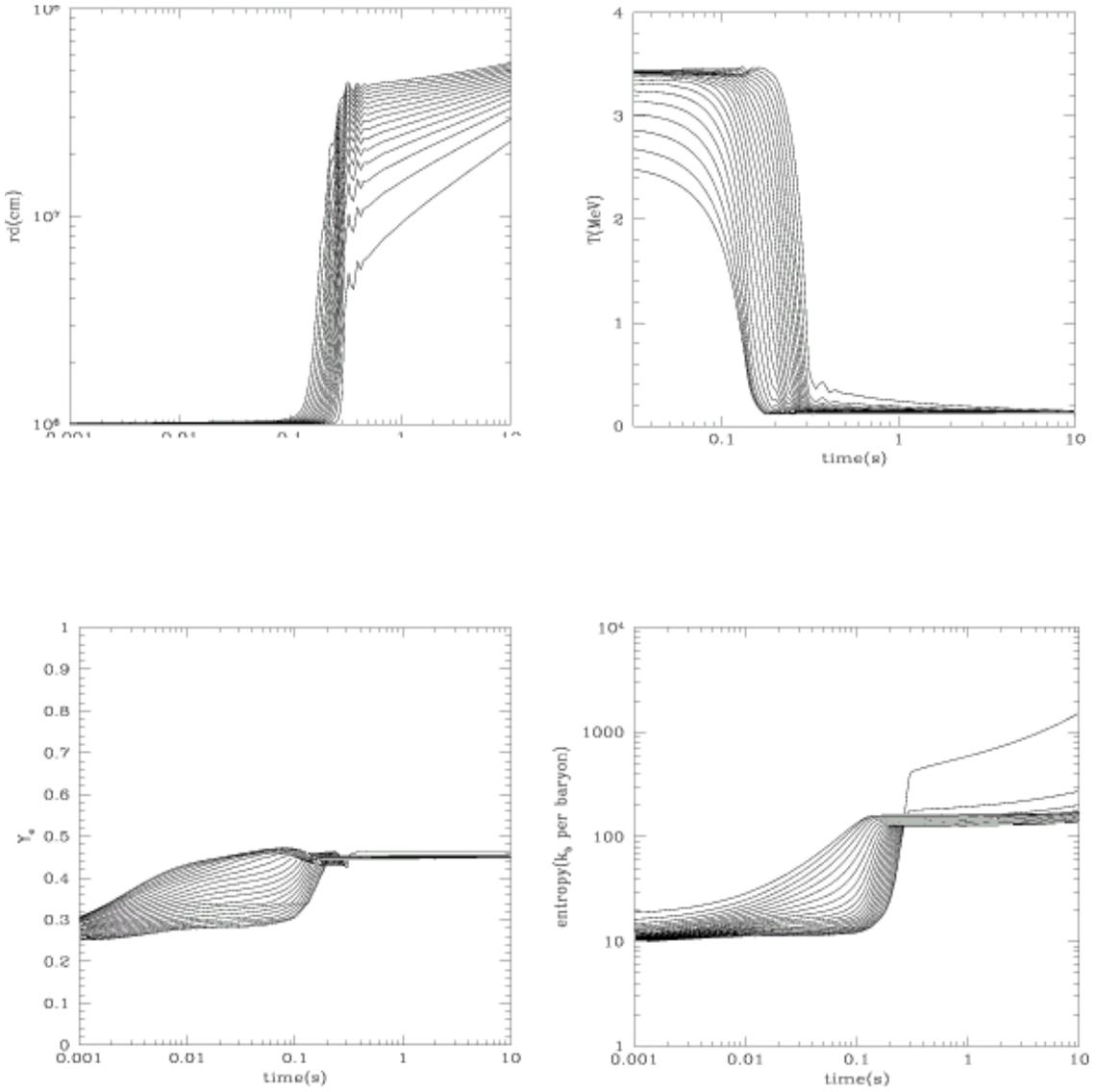}}
  \caption{Trajectories and evolutions of temperature, electron fraction,
   and entropy per baryon for model N. Every 5  mesh points
   are plotted.}
\label{Ntraj}
  \end{center}
\end{figure}

\begin{figure}[h]
 \begin{center}
  \centerline{\includegraphics[width= 15 cm,height=14 cm]{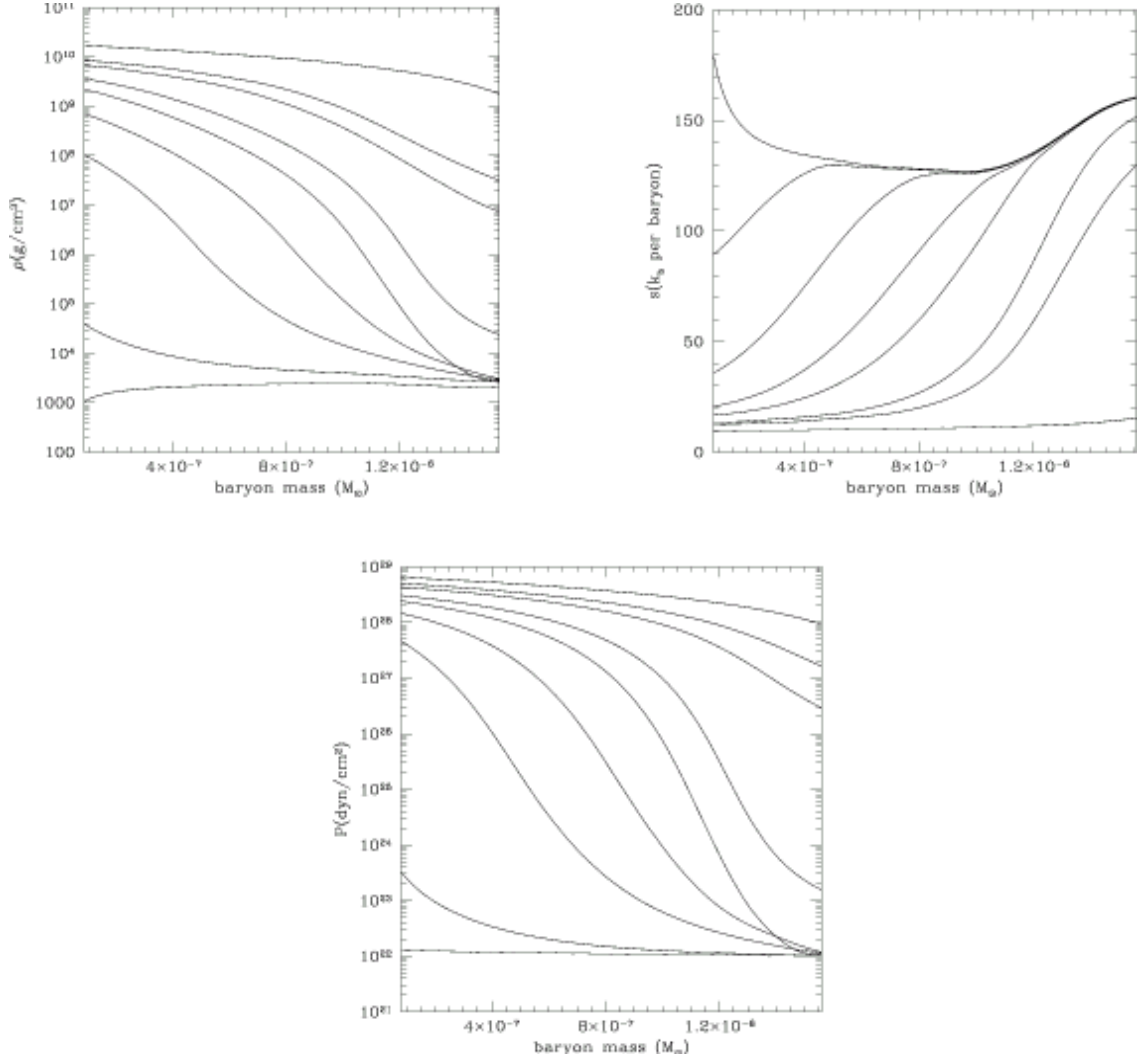}}  \end{center}
 \caption{The baryon mass density $\rho$, 
 pressure $P$, and  entropy $s$ in  model N as a function of baryon
 mass.  Each line
 shows the distribution at t=$2.53\times 10^{-4}$s, $1.00\times 10^{-1}$s, $1.22\times 10^{-1}$s, $1.58\times 10^{-1}$s, $1.80\times 10^{-1}$s, $2.12\times 10^{-1}$s,  $2.48\times 10^{-1}$s, $3.26\times 10^{-1}$s, and $5.00\times 10$s. 
In  the innermost region in the above figures, $\rho$ and $P$ decrease
 and $s$ increases monotonically.  The distribution of $s$ at
 t=$5.00\times10$s is not displayed, since it shows a  false increase
 (see the text for detail). } 
 \label{nonmsev}
\end{figure}

 Figures \ref{Ntraj}-\ref{nonmsev} show the evolution of
 the non-magnetic model (model N). The r-process in the neutrino-driven
 wind proceeds as follows.
 At the surface of the PNS, matter is initially composed of relativistic
 particles, free nucleons, and photons. 
 The  nascent PNS
 contracts and releases $\sim 10^{53}$ergs of gravitational energy
 mostly through neutrino production and emission 
 over its Kelvin-Helmholtz cooling time ($\sim
 10-20$seconds), and a small fraction of the energy is deposited at
 the surface layer via following reactions.   
\begin{equation}
\nu_{e} + n \leftrightarrow e^{-} + p
\label{nu1}
\end{equation}
\begin{equation}
\bar{\nu}_{e} + p \leftrightarrow e^{+} + n
\label{nu2}
\end{equation} 
\begin{equation}
\nu_{i} + e^{-} \leftrightarrow  \nu_{i} + e^{-}
\label{nu3}
\end{equation}
\begin{equation}
\nu_{i} + e^{+} \leftrightarrow  \nu_{i} + e^{+}
\label{nu4}
\end{equation}
\begin{equation}
\bar{\nu}_{i} + e^{-} \leftrightarrow  \bar{\nu_{i}} + e^{-}
\label{nu5}
\end{equation}
\begin{equation}
\bar{\nu}_{i} + e^{+} \leftrightarrow  \bar{\nu_{i}} + e^{+}
\label{nu6}
\end{equation}
\begin{equation}
{\nu}_{i} +   \bar{\nu_{i}}  \leftrightarrow  e^{-} + e^{+}
\label{nu7}
\end{equation}
 Here the subscript $i$ denotes the flavor of neutrino ($i =
 e,\mu,\tau$). Most efficient reactions to deposit energy are 
 the neutrino absorption on free
 nucleons  ((\ref{nu1}) and  (\ref{nu2})).  These reactions also determine
 the evolution of the electron fraction.
 When sufficient energy is deposited 
 to  overcome the gravitational attraction, an outflow (neutrino-driven wind)
 is generated. The wind is heated until it reaches $\sim$ 50km.
 As the wind expands, the temperature decreases and when it drops
 below $\sim 1$MeV, the free nucleons start to combine with each other and form
 $\alpha$-particles. At this stage, neutrino reactions have ceased, and
 the subsequent evolution becomes adiabatic. When the temperature drops
 below $\sim 0.5$MeV, $\alpha$-process occurs (e.g. triple $\alpha$
 reactions), and seed nuclei of the r-process are formed 
 until the temperature drops to $\sim 0.2$MeV.

 There are three key quantities in the wind scenario which determine
 the neutron-to-seed ratio. Those are the electron fraction ($Y_{e}$)
 and entropy per baryon ($s$), and  
 the dynamical time scale ($\tau_{dyn}$).
 The  dynamical timescale is defined to be the duration of the 
 $\alpha$-process. 
 For a robust r-process, lower values are favored for $Y_e$ and
 $\tau_{dyn}$ because it implies that  neutrons are rich and the duration
 of the production of seed nuclei  is short. Higher entropy is favored for
 the following reason.
 When the system is still in high temperature
 regime  (above $\sim 0.5$MeV), and nuclear statistical equilibrium
 (NSE)
 is maintained, the
 high entropy shifts the equilibrium to the state in which 
 matter is composed mainly of $\alpha-$particles and free neutrons with
 only a small fraction of nuclei.
 However, when the temperature becomes  $\lesssim0.5$MeV,
 the system falls out of NSE. The high entropy implies a low
 density. Therefore, the $\alpha$-process proceeds slowly, and
 most of the matter remains  as $\alpha$-particles.

 In  model N, the wind expands rapidly  $\sim$0.1 s after the beginning of
 neutrino heating processes. As the 
 temperature decreases,  the $\alpha$-processes occur,
 which corresponds to the location from $\sim$50km to $\sim$150km.
 The 
 state is reached to the asymptotic one 
 with a constant temperature $\sim$0.1MeV  at $\sim$0.2s.
 Rapid
 increases of  the electron fraction and entropy are seen when the wind is
 expanding  from the surface of the PNS to $\sim$50km.
 Note that a slight increase of  entropy in the late
 phase of the wind evolution
 is caused by a breakdown of approximation for the heating
 by electron and positron scatterings.
 This occurs because the equation we have  used for the heating due
 to  electron and positron scatterings is valid only when the
 temperature is high 
 and, therefore, overestimates the increase of the entropy in the low
 temperature regime  (see Ref.~\cite{rf:8} for detail). 
 This takes place in all the models commonly.
 However, this  affects little
 the dynamics of the wind in this phase,
 and does not make a significant change
 in the dynamical time scale, electron fraction
 and  entropy before the $\alpha$-process stage. Thus, we can safely discuss
 the possibility of r-process  by ignoring  this false increase in 
 entropy. 
 Figure \ref{nonmsev} shows  
 the baryon mass density
 $\rho$,  matter pressure $P$, and  entropy per baryon $s$ as a 
 function of baryon mass. The density and pressure show
 significant decrease when the wind is expanding rapidly. Then the
 asymptotic state appears where $\rho\sim10^3$g/cm$^3$ and
 $P\sim10^{22}$dyn/cm$^2$ in the whole region.
  These features are identical with the
 previous studies done by Ref.~\cite{rf:8}

\begin{figure}
  \begin{center}
  \centerline{\includegraphics[width= 15 cm,height=15 cm]{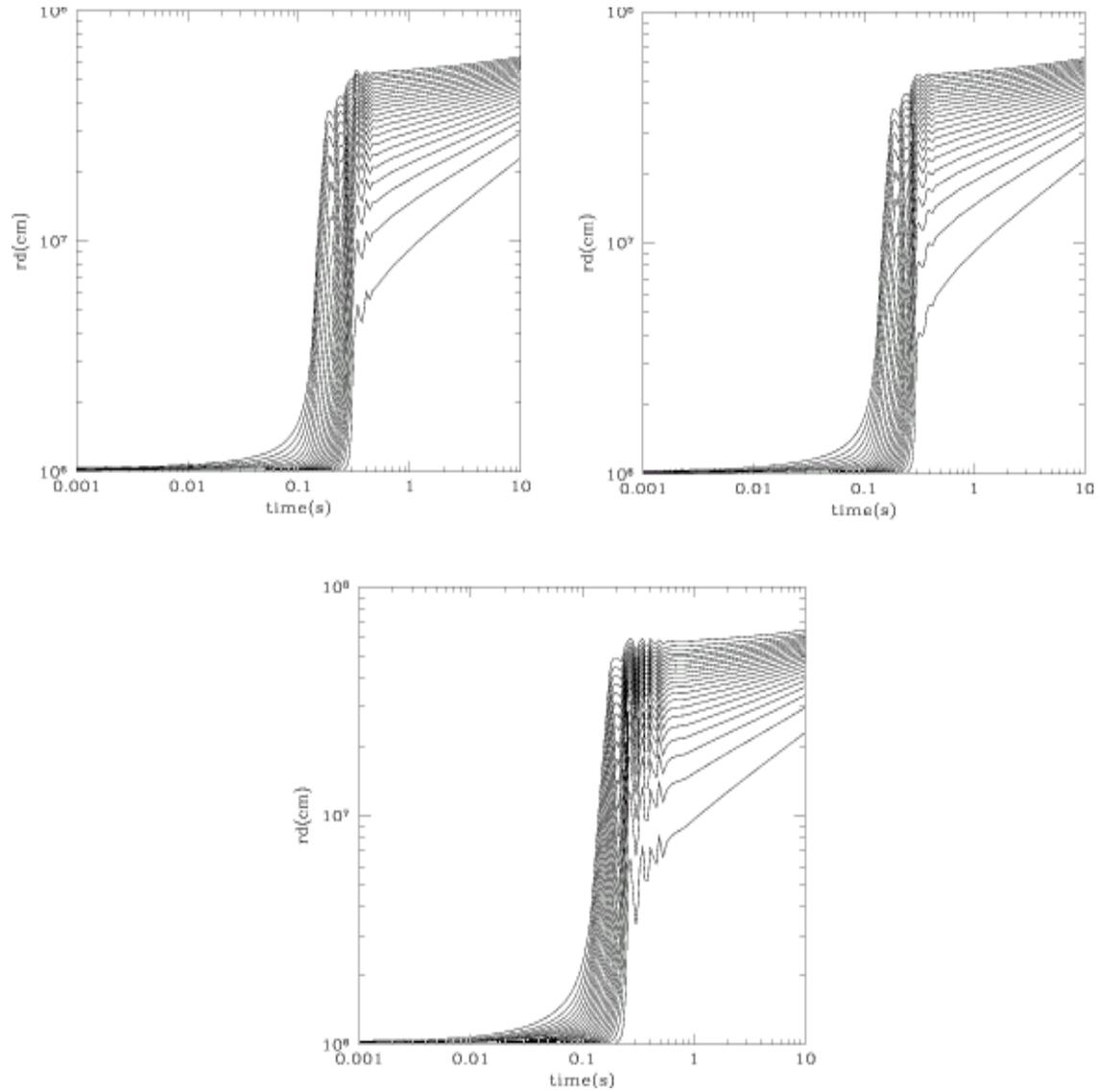}}

  \caption{Trajectories for model $10^6$M (top left panel),
   $\frac{1}{10}$M (top right panel), and $\frac{1}{200}$M (bottom panel). Every 5  mesh points are
   plotted.}
\label{1/200traj}

  \end{center}
\end{figure}

\begin{figure}
\begin{center}

  \centerline{\includegraphics[width= 15 cm,height=15 cm]{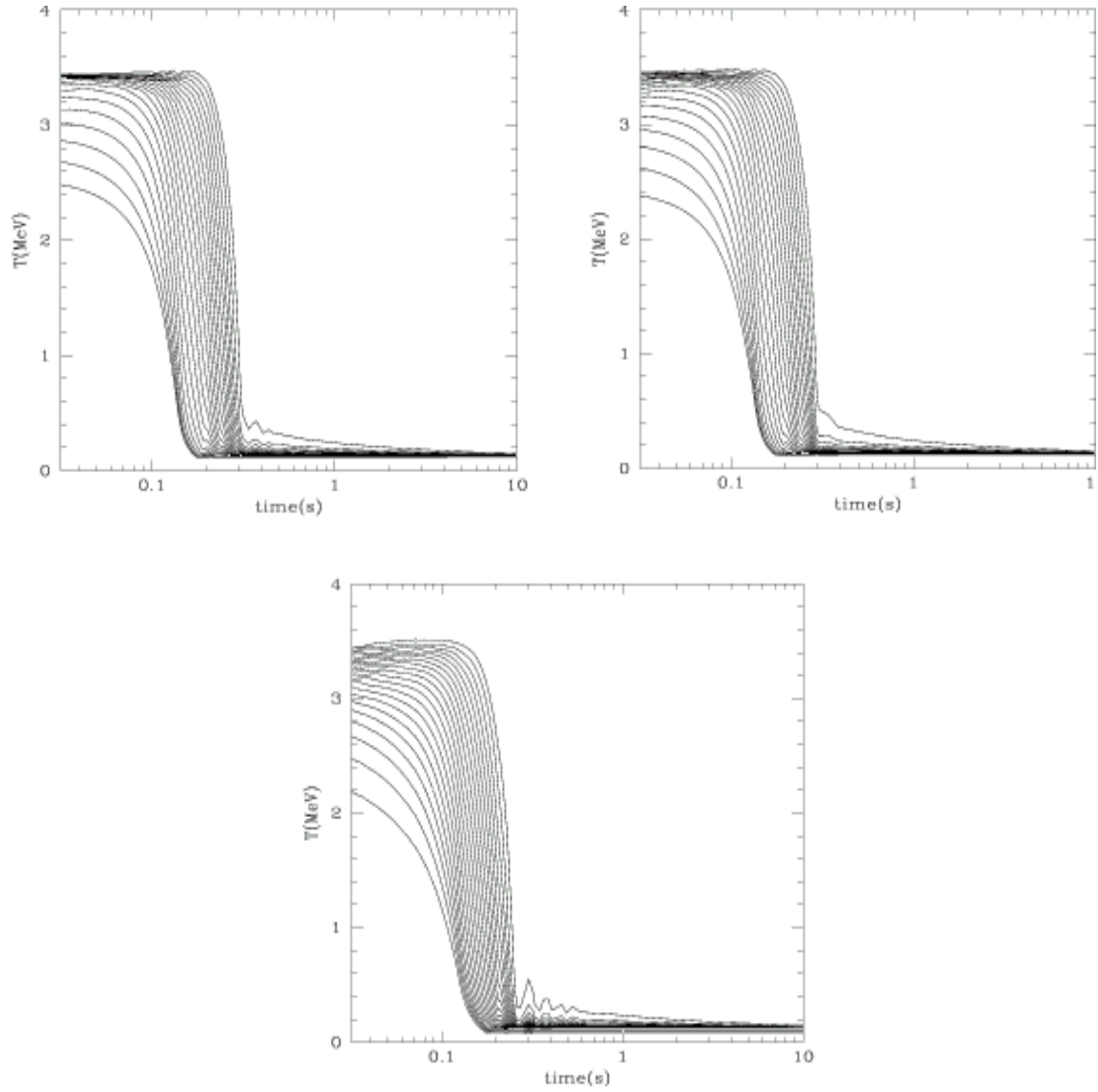}}
  \caption{Evolution of the temperature for  model $10^6$M (top left
 panel), $\frac{1}{10}$M (top right panel), and $\frac{1}{200}$M (bottom
 panel). Every 5 mesh points are plotted.}
\label{200-1temp}

 \end{center}
\end{figure}

\begin{figure}
\begin{center}

  \centerline{\includegraphics[width= 15 cm,height=15 cm]{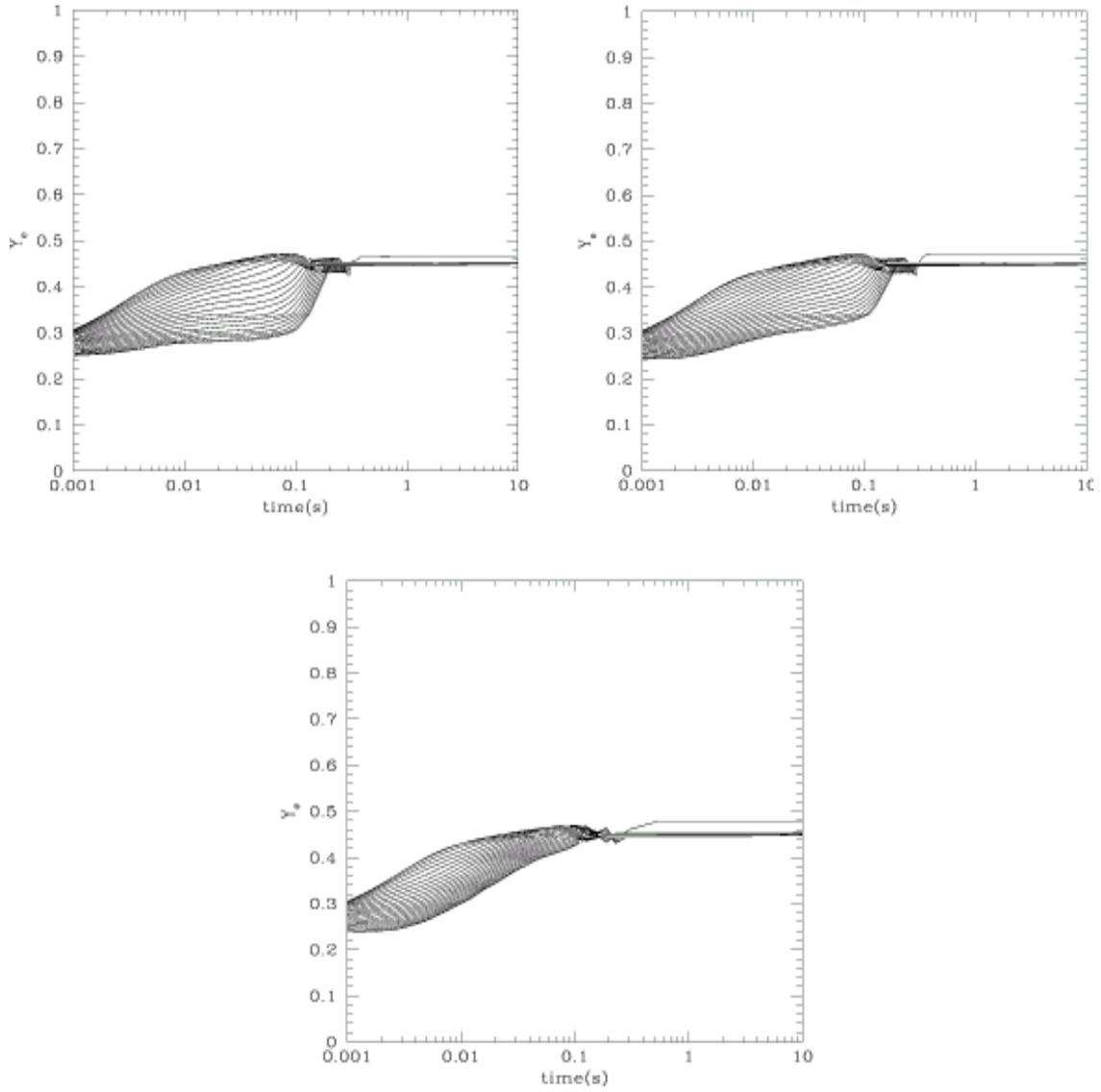}}
  \caption{Evolution of the electron fraction for model $10^6$M (top
 left panel), $\frac{1}{10}$M (top right panel), and  $\frac{1}{200}$M
 (bottom panel). Every 5  mesh points  are plotted.}
\label{200-1Ye}

 \end{center}
\end{figure}

\begin{figure}[h]
  \begin{center}
  \centerline{\includegraphics[width= 11 cm,height=9 cm]{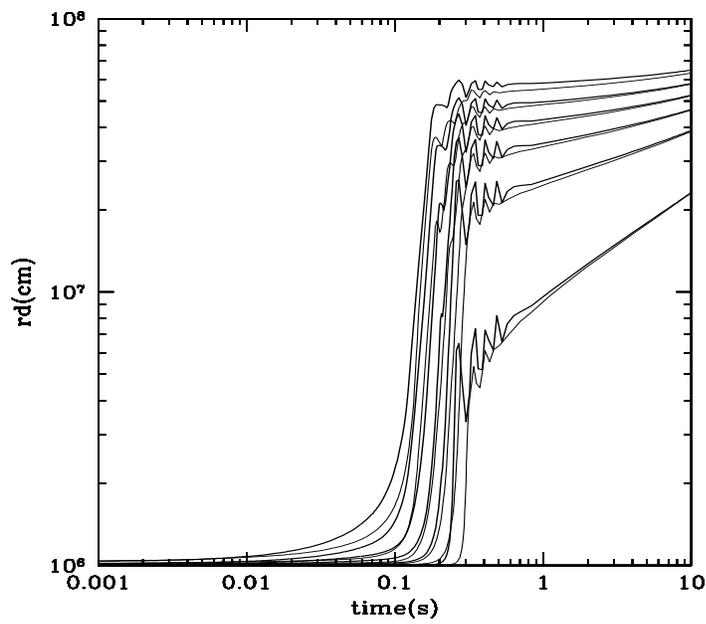}}
  \caption{Trajectories for model $\frac{1}{200}$M(thick solid lines) and
    trajectories for model N(thin solid lines). Every 20 mesh points are plotted. }
\label{non200-1rd}
  \end{center}
\end{figure}

 The trajectories of   models $10^6$M, $\frac{1}{10}$M, and
 $\frac{1}{200}$M are shown in figure \ref{1/200traj}.
  As the
 initial magnetic field becomes greater, more rapid expansion can be seen.
 However, the later evolution shows only little deviation from
  model N (see figure
 \ref{non200-1rd}),
 and the overall evolution for all models is
 nearly identical with that for the non-magnetic model.
 Figures \ref{200-1temp}  and  \ref{200-1Ye}
 present, respectively, the evolutions of temperature  and  electron fraction
 for models $10^6$M, $\frac{1}{10}$M, and
 $\frac{1}{200}$M. 
 As in model N, the efficient decrease in  temperature and increase in
 electron fraction   take   place when the
 matter is rapidly expanding. 
 The values of  temperatures and electron fractions at the same
 radius almost coincide among all the models including model N.
 This implies that the models with stronger magnetic fields
 show an earlier drop in  temperature and  rise in the electron
 fraction, and  the later behavior becomes identical. 
 The asymptotic values of $Y_{e}$ and $T$ do not vary so much  among all models
 ($Y_e \sim$0.45,  $T\sim 0.1$MeV).

 The distributions of 
 the baryon mass density $\rho$,  entropy per baryon $s$,  
 matter pressure $P$ as well as magnetic pressure $B^2/8\pi$  for these three
 models are given  (as a function of the baryon mass) in
 figures \ref{10^6msev}-\ref{200-1msev}.  
 The evolutions of  baryon mass density and pressure are essentially 
 same in all models (including model N). 
 The density and pressure  decrease as the wind expands and reach an
  asymptotic value, which is nearly identical among all models  
 ($\rho \sim 10^3$g/cm$^3$, $p \sim 10^{22}$dyn/cm$^2$). 
 The evolutions of the magnetic pressure and
 entropy
 are also similar among the models
 except for the asymptotic value. 
 As the
 wind expands, the initially homogeneous magnetic pressure decreases and
 reaches  an
 asymptotic value. Unlike the density and pressure, the asymptotic value
 varies largely in the region. The final distribution of magnetic
 pressure
 has a positive gradient, and the outer region is  $\sim$2 orders of
 magnitude larger
 than the inner region.  
 Since the neutrino heating processes occur up to $r\sim$50km, the
 entropy increases until the wind reaches that radius.  
 The models with initially strong magnetic fields tend to have lower
 asymptotic values compared to the  models with weaker fields.
 This tendency can be
 observed particularly in the outer region of the wind.
 As mentioned above, the more rapid evolutions of these values ($\rho$,
 $P$, $s$, and $B^2/8\pi$) seen in the models with stronger magnetic
 fields are results of the more rapid expansion.
 These
 behaviors  will be discussed in more detail in the next  section.

 The three key quantities ($s$, $\tau_{dyn}$, and $Y_e$) for the r-process
 are summarized for all the models in Table \ref{table:2}. 
 Since the outer and inner regions  are  affected 
 by the boundary conditions,
 the average values shown in the table are obtained by
 discarding the inner and  outer 5  mesh points. The  entropy in the table
 is the average value at the beginning of the $\alpha$-process
 (T$\sim$0.5MeV).
 The  false increase of entropy caused by the
 breakdown of the approximation in the late phase is ignored.
 The dynamical time scale $\tau_{dyn}$ is calculated as the
 duration, in which the temperature  decreases  from 0.5MeV to 0.2MeV,
 for each trajectory and averaged.  
 As noted above, 
 we can not see any
 significant difference  in the trajectories
 of the  magnetic wind cases and  non-magnetic wind case. Even in the models 
 with  magnetar-scale field strengths, only a slight change comes out
 (the trajectories of model $\frac{1}{200}$M together
 with model N are displayed in figure \ref{non200-1rd}).
 Therefore, it is no surprise that the
 evolution of the variables such as temperature, electron fraction, 
 density, pressure and
 entropy are also nearly identical (see figures \ref{Ntraj}-\ref{nonmsev}
 and \ref{200-1temp}-\ref{200-1msev}). As a result, even strong magnetic
 fields give remarkably little change in the three key parameters
 (see Table \ref{table:2}). We will discuss the reason for this in the
 next section in greater detail.

\begin{table}[h]
 \begin{center}
  \begin{tabular}{ccccc}
 \hline
 \hline
  Model  & $s(\textstyle{k_b})$ &  $\tau_{dyn}(\textstyle{s})$  & $Y_{e}$  \\[2pt]

 \hline
         &&&&\\[-10pt]

    N    &   137   & 4.63$\times$10$^{-2}$   &   0.453\\[3pt]

 10$^6$M & 137  &  4.57$\times$10$^{-2}$ & 0.453 \\[3pt]

 10$^3$M & 136  & 4.57$\times$10$^{-2}$ & 0.453 \\[3pt]

 10M  & 136   & 4.55$\times$10$^{-2}$  & 0.453 \\[3pt]

  1M  &  137  & 4.55$\times$10$^{-2}$  & 0.453 \\[3pt]

 $\frac{\textstyle{1}}{\textstyle{10}}$M  & 136    &  4.41$\times$10$^{-2}$   & 0.453 \\[3pt]

  $\frac{\textstyle{1}}{\textstyle{20}}$M  & 135    & 5.27$\times$10$^{-2}$  &  0.453 \\[3pt]

  $\frac{\textstyle{1}}{\textstyle{50}}$M  &  133    & 5.35$\times$10$^{-2}$  & 0.453 \\[3pt]

  $\frac{\textstyle{1}}{\textstyle{100}}$M  & 129     & 4.01$\times$10$^{-2}$  & 0.453 \\[3pt]

  $\frac{\textstyle{1}}{\textstyle{200}}$M  &  128   & 5.45$\times$10$^{-2}$  & 0.453 \\[3pt]

 \hline  
  \end{tabular}
 \end{center}
\caption{Three key quantities for each model. These values are
 the average over the whole wind after  discarding the inner  and 
 outer 5  mesh points.} 
\label{table:2} 
\end{table}

\section{Discussions}
 Our results show that the homogenous magnetic field has only a
 little effect on the wind dynamics. In particular,  the three key
 quantities for the r-process  do not vary much from those for
 the non-magnetic wind.
 This is true even for the very strong magnetic fields.
 In this section, we will analyze  this behavior in more detail.

\begin{figure}
  \begin{center}
  \centerline{\includegraphics[width= 11 cm,height=9 cm]{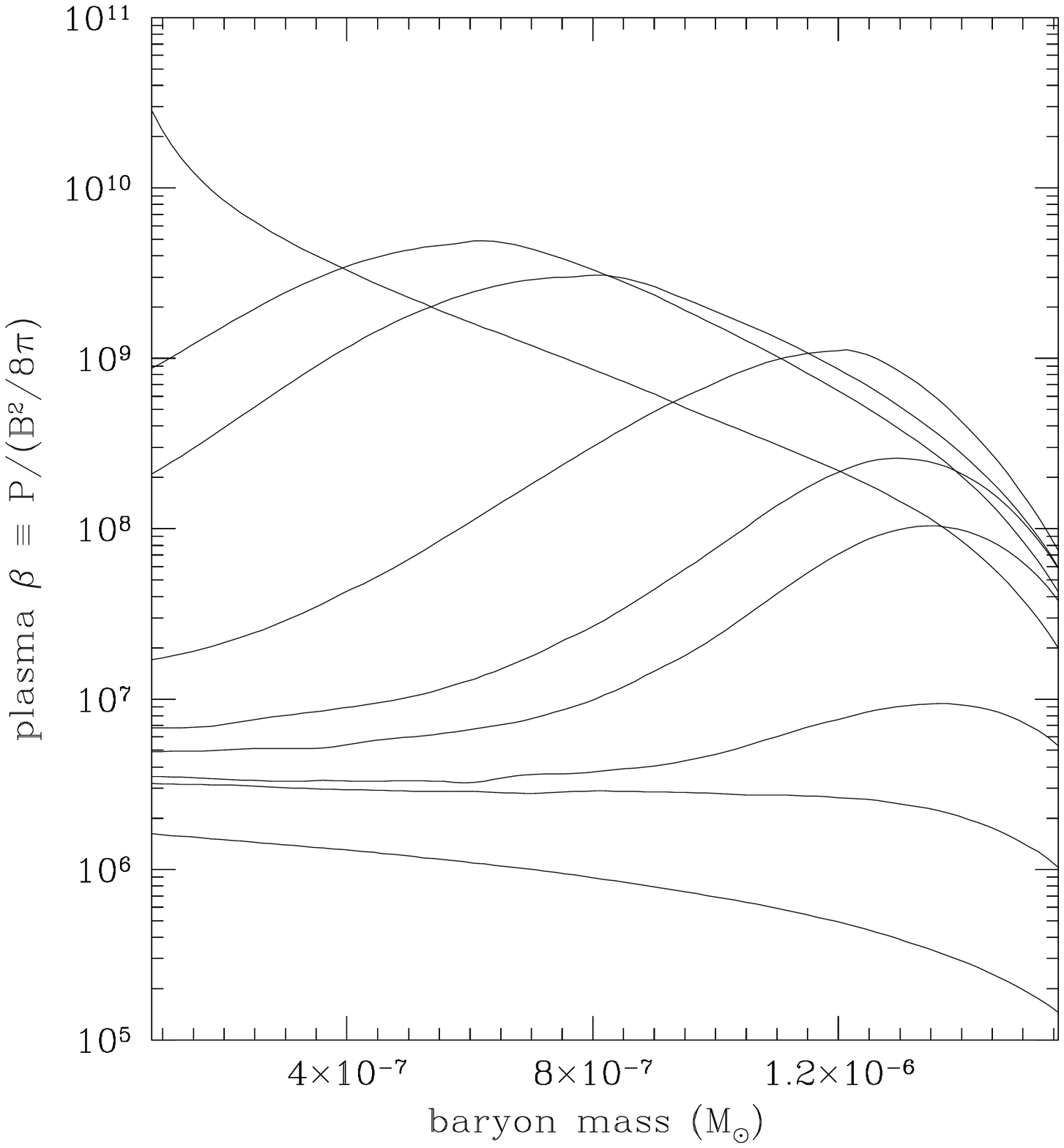}}
  \caption{Evolution of the plasma beta for model $10^6$M. 
     Each line
 shows the distribution at t=$2.53\times 10^{-4}$s, $1.49\times 10^{-2}$s, $4.12\times 10^{-2}$s, $1.00\times 10^{-1}$s, $1.22\times 10^{-1}$s, $1.58\times 10^{-1}$s, $2.12\times 10^{-1}$s, $2.33\times 10^{-1}$s, and $5.00\times 10$s.
 In  the innermost region in the above figure, $\beta$ increases monotonically in time.  } 
\label{10^6beta}
  \centerline{\includegraphics[width= 11 cm,height=9 cm]{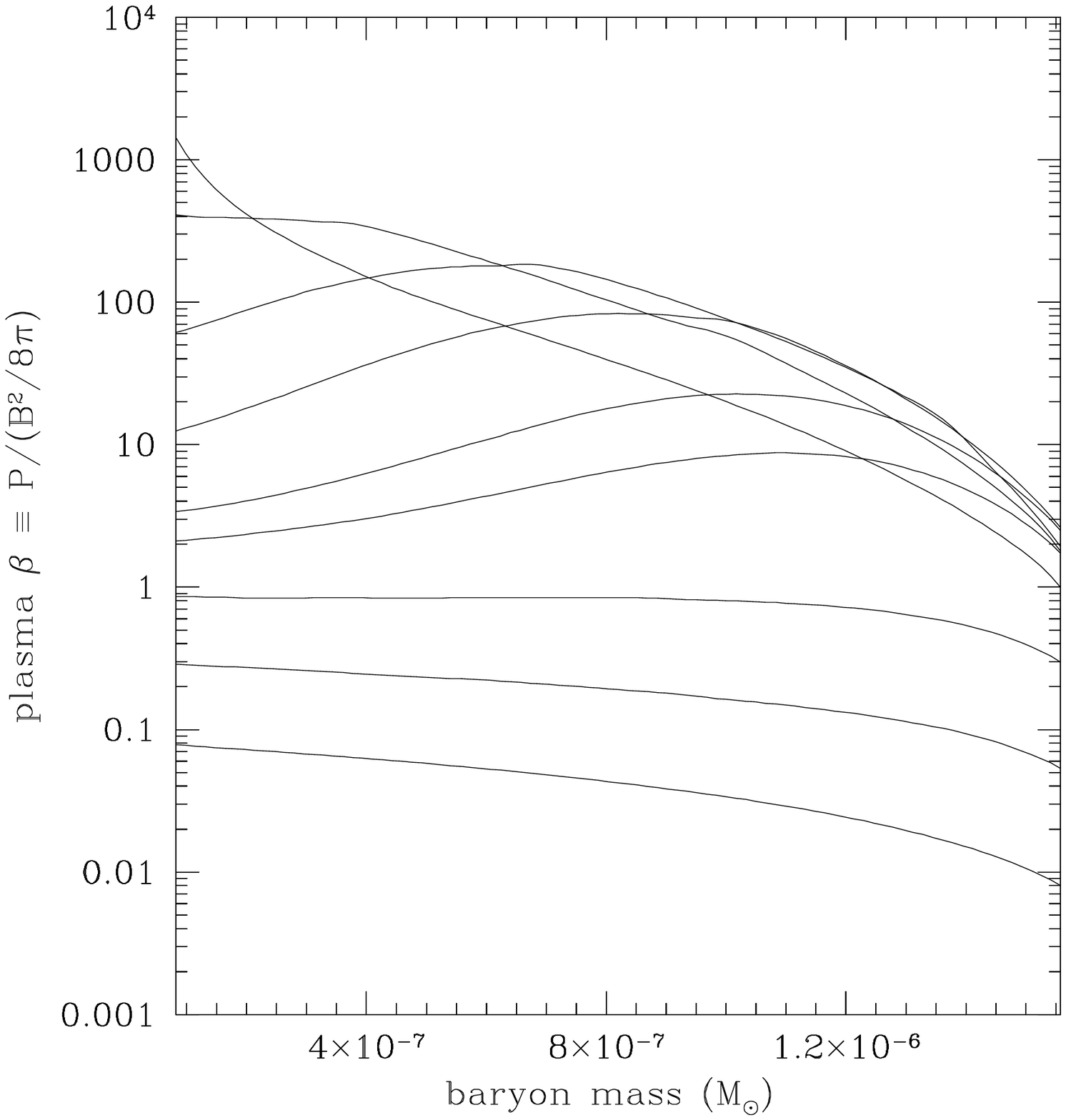}}
  \caption{Evolution of the plasma beta for model $\frac{1}{200}$M.
 Each line
 shows the distribution at t=$7.24\times 10^{-5}$s, $1.41\times 10^{-2}$s, $4.10\times 10^{-2}$s, $1.00\times 10^{-1}$s, $1.22\times 10^{-1}$s, $1.56\times 10^{-1}$s, $1.83\times 10^{-1}$s, $2.13\times 10^{-1}$s, and $5.00\times 10$s.
 In the innermost region in the above figure, $\beta$ increases
   monotonically in time. }
\label{200-1beta}
  \end{center}
\end{figure}

 The reason why the dynamics is little affected by magnetic fields is
 understood from the evolutions of  magnetic fields.
 Figures \ref{10^6beta}-\ref{200-1beta}
 show the evolutions of the plasma beta, which is defined to be the
 ratio of matter pressure to magnetic pressure, for models $10^6$M
 and $\frac{1}{200}$M. In  these models, the plasma beta
 increases rapidly from the initial values and the magnetic pressure
 becomes negligible for the wind dynamics.
 This tendency can be explained analytically as follows.

 From equation (\ref{Bs}), the variation rate of the magnetic pressure is 
 given as
\begin{equation}
\frac{1}{B_{\theta}^2}\frac{d B_{\theta}^2}{d t} = -2\frac{v}{r} -2\frac{d \lambda}{d t} , 
\label{dtdB}
\end{equation}
 where $d/dt$ denotes the Lagrange derivative, and $v$ is a radial fluid
 velocity which is defined as
\begin{equation}
 v \equiv \frac{dr}{dt}  .
\label{v}
\end{equation} 
 Ignoring the neutrino heating, which is important only at the earliest
 stage, we can evaluate the variation
 rate of matter pressure  approximately  as
\begin{equation}
\frac{1}{P} \frac{d P}{d t}   = \gamma \frac{1}{\rho}\frac{d \rho}{dt } , 
\label{adiabatic}
\end{equation} 
\begin{equation}
 \gamma \equiv \left(\frac{\partial \ln P}{\partial \ln\rho}\right)_{s} ,
\label{adindex}
\end{equation}
 where $\gamma$  is the adiabatic index. 
 Ignoring further general relativistic corrections,
 $e^{\lambda}$ can be approximated as 
\begin{equation}
 e^{\lambda} \sim \left(\frac{\partial m}{\partial r} \right)^{-1} \sim \frac{1}{4\pi r^2 \rho} .
\label{lambda2}
\end{equation}
 From the above equation, $\displaystyle \frac{d\lambda}{dt}$ is written as
\begin{equation}
 \frac{d\lambda}{dt}  \sim -\frac{1}{\rho}\frac{d\rho}{dt} - 2\frac{v}{r} .
\label{lambda3}
\end{equation}
 With the use of the continuity equation and equation (\ref{lambda3}),
 equations (\ref{dtdB}) and (\ref{adiabatic}) become, respectively
\begin{equation}
\frac{1}{B_{\theta}^2}\frac{d B_{\theta}^2}{d t} = 2\frac{v}{r} -2\left(\frac{\partial v}{\partial r}\right)_{t} ,  
\label{dtdB2}
\end{equation}
\begin{equation}
\frac{1}{P}\frac{dP}{dt} = -2\gamma \frac{v}{r} -\gamma \left(\frac{\partial v}{\partial r} \right)_{t}  . 
\label{adiabatic1-2}
\end{equation}
 If we locally approximate the velocity profile by
 a power-law, $v \propto r^{\alpha}$,
 the variation rates are evaluated  as
\begin{equation}
\frac{1}{B_{\theta}^2}\frac{d B_{\theta}^2}{d t} \propto -(2 + 2\alpha)r^{\alpha -1}  ,
\label{dtdB3}
\end{equation}
\begin{equation}
\frac{1}{P}\frac{dP}{dt}  \propto -(2 + \alpha)\gamma r^{\alpha -1}  .
\label{adiabatic2}
\end{equation}
 Since the adiabatic index  $\gamma$ is approximately 4/3,  we finally
 obtain the criterion  that if   $\alpha<1$, the plasma beta is decreased.

\begin{figure}
  \begin{center}
  \centerline{\includegraphics[width= 11 cm,height=9 cm]{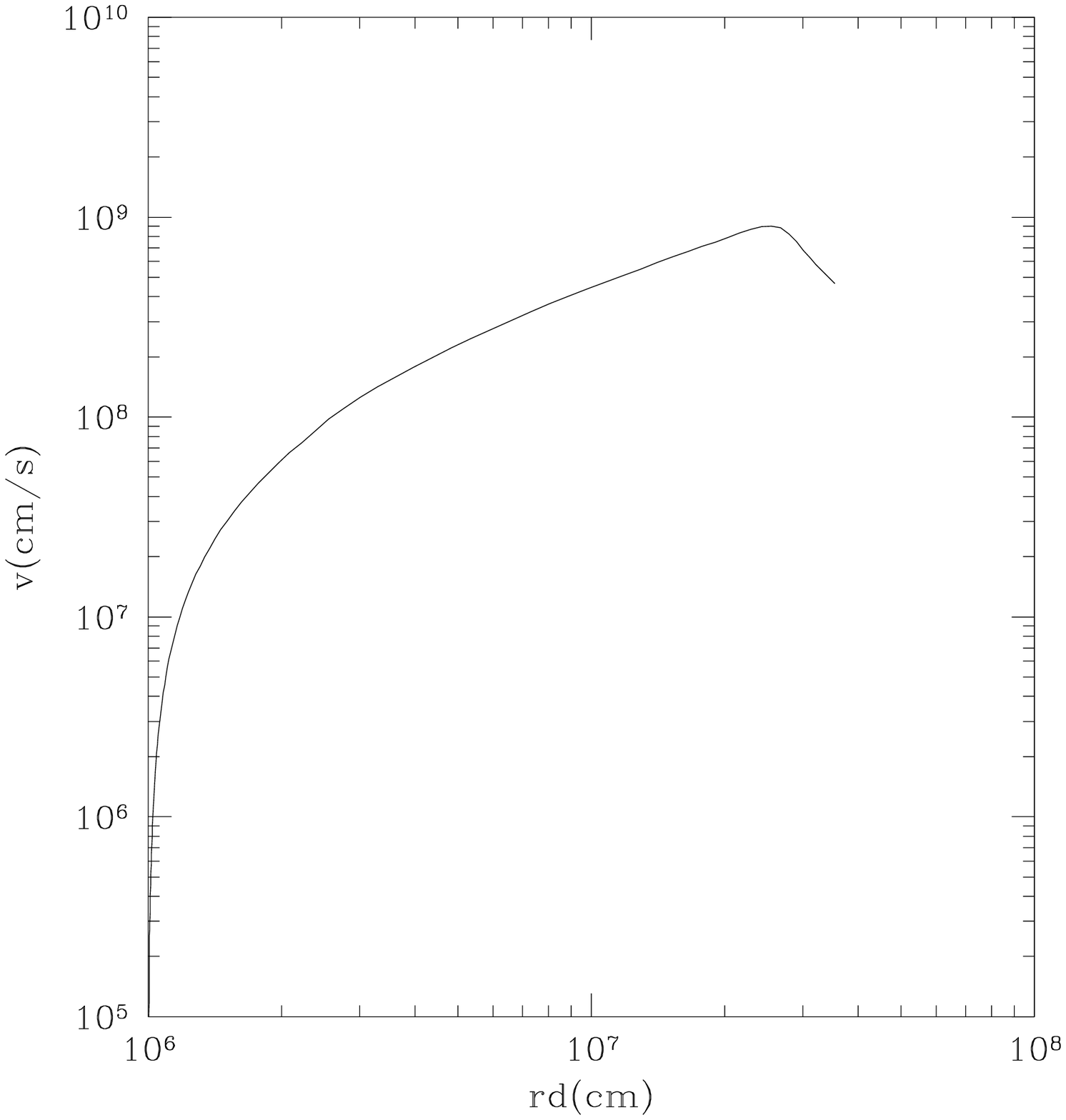}}
  \caption{Velocity profile for model N at t = $2.33\times 10^{-1}$s.  }

\label{rdv}
  \end{center}
\end{figure}

 Figure \ref{rdv}
 shows the radial profile of velocity for model N 
 when the wind has settled to a quasi-steady state.
 Though this is a snapshot for this model,  all the models have a similar
 configuration. If we apply the above criterion in this figure, we find
 that the region in which the plasma beta should decrease 
 is the outer region ($r \geq 60$km). It is noted that the assumptions
 we have made to derive the above condition, that is, the adiabaticity and
 nonrelativisity of the flows, are clearly satisfied in the region.
 Thus, after the wind expands beyond $\sim$60km, the plasma 
 beta will 
 decrease.
 Since we are considering the region exterior to the PNS surface, the
 Newtonian gravity  is not so bad an approximation in all region, and
 equation (\ref{dtdB3}) is  valid even in the inner region ($r<60$km).
 On the other hand, equation (\ref{adiabatic2}) is not appropriate in
 this  inner region, since the heating cannot be ignored and
 the adiabatic condition is not satisfied. However,
 it is obvious that the heating processes tend to increase the matter
 pressure and, as a result, 
 increase the plasma beta more than in the
 adiabatic flow.
 Thus, although the obtained criterion is not valid,
 the plasma beta is expected to increase in the inner region where
 $\alpha > 1$.

 From the above discussions, the wind is expected to expand initially with
 increasing plasma beta up to $r\sim$60km, and thereafter
 with decreasing plasma beta.
 Our numerical results actually agree with the expectation. This can
 be confirmed by seeing figures
 \ref{10^6beta}-\ref{200-1beta}.
 Since the increase
 of the plasma beta occurs more rapidly than the following decrease,
 the wind tends to be matter-pressure-dominant in the entire evolution.
 Thus, even models with very 
 strong magnetic fields
 give only a little effect on the wind evolution.
 In the following, we  will further analyze, although small,
 effects of magnetic fields on the dynamics, focusing on
 model $\frac{1}{200}$M.

\begin{figure}[h]
 \begin{center}
  \centerline{\includegraphics[width= 15 cm,height=14 cm]{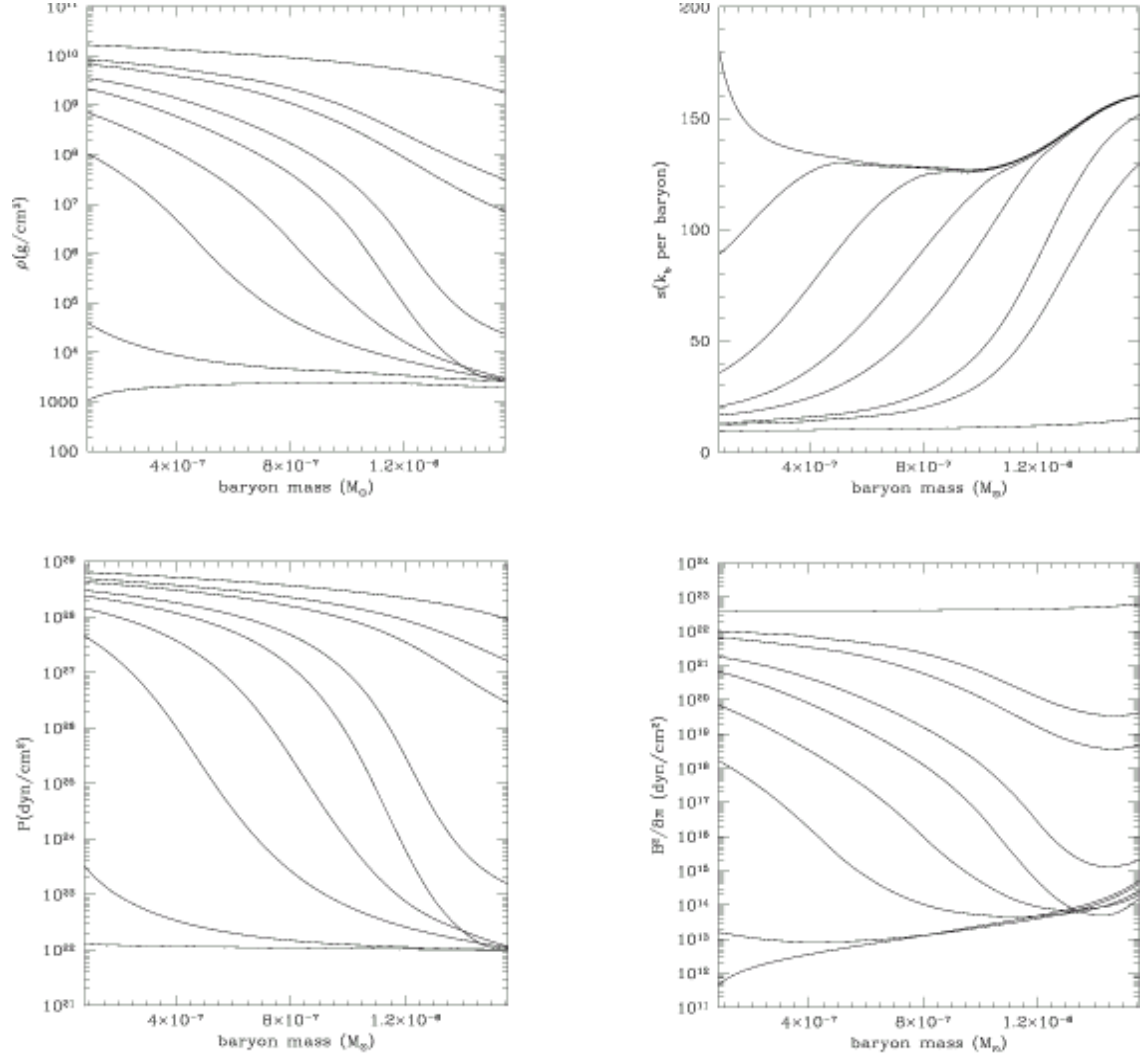}}  \end{center}
 \caption{The baryon mass density $\rho$, 
 pressure $P$,  entropy $s$, and  magnetic pressure
 $B^2/8\pi$ for 
 model $10^6$M as a function of baryon mass.  Each line
 shows the distribution at  t=$2.53\times 10^{-4}$s, $1.00\times 10^{-1}$s, $1.22\times 10^{-1}$s, $1.58\times 10^{-1}$s, $1.80\times 10^{-1}$s, $2.12\times 10^{-1}$s, $2.48\times 10^{-1}$s, $3.26\times 10^{-1}$s, and $5.00\times 10$s.
In the innermost region in the above figures, $\rho$, $P$, and
 $B^2/8\pi$ decrease and $s$ increases monotonically.  The distribution
 of $s$ at t=$5.00\times10$s is not displayed, since it shows a later
 false increase (see the text for detail).} 
 \label{10^6msev}
\end{figure}

\begin{figure}[h]
 \begin{center}
  \centerline{\includegraphics[width= 15 cm,height=14 cm]{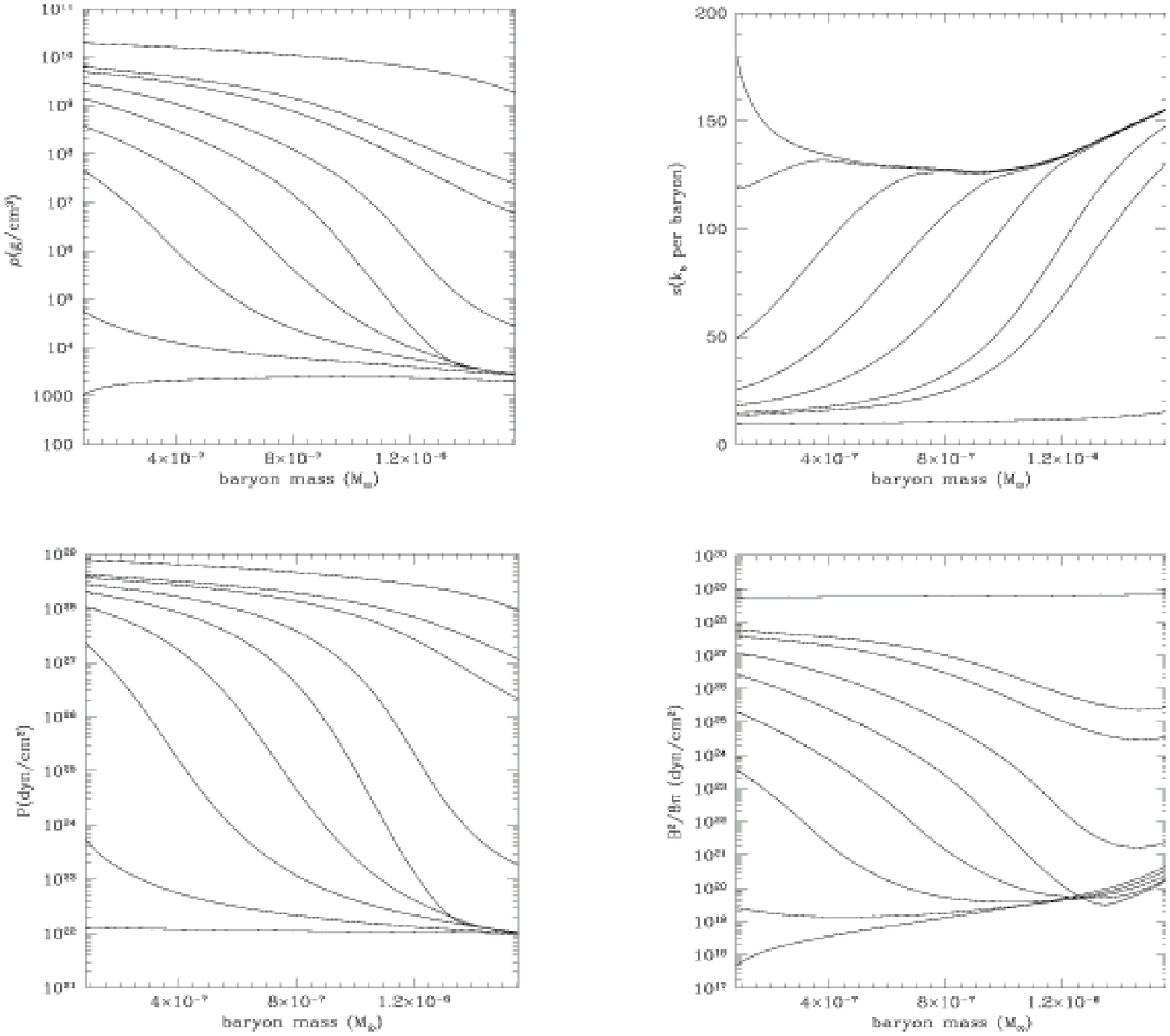}}  \end{center}
 \caption{The baryon mass density $\rho$, 
 pressure $P$, entropy $s$, and  magnetic pressure
 $B^2/8\pi$ for 
 model $\frac{1}{10}$M as a function of baryon mass.  Each line
 shows the distribution at  t=$4.26\times 10^{-5}$s, $1.02\times 10^{-1}$s, $1.22\times 10^{-1}$s, $1.56\times 10^{-1}$s, $1.85\times 10^{-1}$s, $2.16\times 10^{-1}$s, $2.53\times 10^{-1}$s, $3.40\times 10^{-1}$s, and $5.00\times 10$s.
 In  the innermost region in the above figures, $\rho$, $P$, and
 $B^2/8\pi$ decrease and $s$ increases monotonically. The distribution
 of $s$ at t=$5.00\times10$s is not displayed, since it shows a  false
 increase (see the text for detail).} 
 \label{10-1msev}
\end{figure}

\begin{figure}[h]
 \begin{center}
  \centerline{\includegraphics[width= 15 cm,height=14 cm]{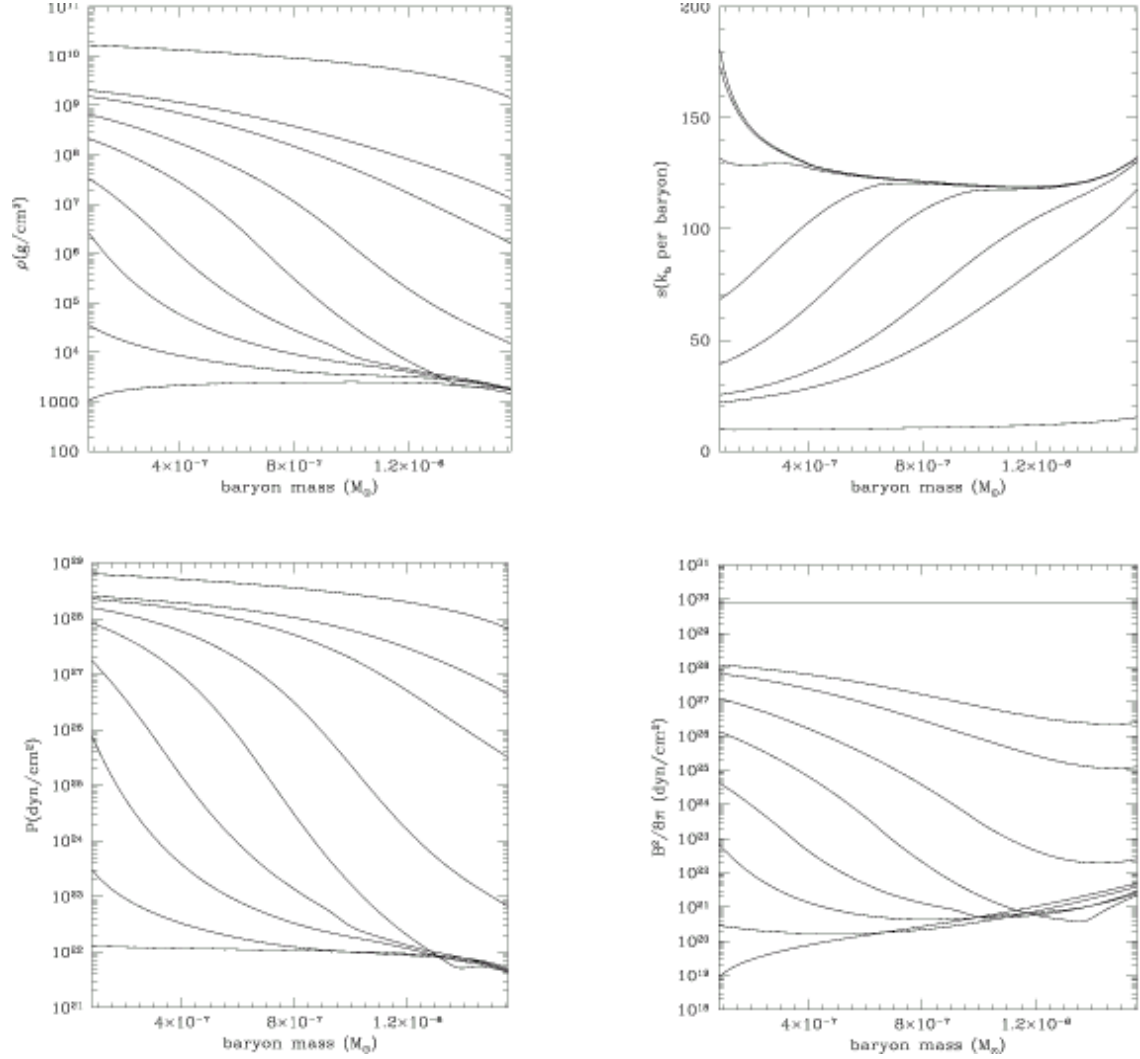}}  \end{center}
 \caption{The baryon mass density $\rho$, 
 pressure $P$,  pressure $s$, and  magnetic pressure
 $B^2/8\pi$ for 
 model $\frac{1}{200}$M as a function of baryon mass mesh.  Each line
 shows the distribution at t=$7.24\times 10^{-5}$s, $1.00\times 10^{-1}$s, $1.22\times 10^{-1}$s, $1.56\times 10^{-1}$s, $1.83\times 10^{-1}$s, $2.13\times 10^{-1}$s, $2.34\times 10^{-1}$s, $2.84\times 10^{-1}$s, and $5.00\times 10$s.
In the innermost region in the above figures, $\rho$, $P$, and
 $B^2/8\pi$ decrease and $s$ increases monotonically. The distribution
 at t=$5.00\times10$s is not displayed, since it shows a  false increase
 (see the text for detail).} 
 \label{200-1msev}
\end{figure}

 At the beginning of expansion,
 the magnetic pressure is  dominant, as can be seen
 from figures \ref{10^6msev}-\ref{200-1msev}, and its negative
 gradient pushes the matter outward. This causes an early rapid expansion
 seen in figure \ref{non200-1rd}. Although the magnetic
 pressure accelerates  matter for a while,
 the effects of the magnetic pressure soon 
 becomes significantly small (since the plasma beta decreases rapidly).
 Then, as they approach the asymptotic state, the gradient of the
 magnetic field changes signs and starts to decelerate matter (see
 figures \ref{10^6beta}-\ref{200-1beta}). 
 Note that although magnetic pressure is  small at this stage,
 it still has some leverage on the dynamics  to cancel 
 the initial accelerations. This causes the later catch-up
 of non-magnetic wind model N seen in
 figure \ref{non200-1rd}.

 The  subsequent distribution of the magnetic pressure  is a direct
 consequence of
 the  initial assumption on the magnetic field. The initial  homogeneity
 of 
 magnetic fields implies that 
 the matter near the surface
 has a greater magnetic flux per unit mass  compared with the matter
 deeper inside, since the density is higher there. 
 As the wind evolves to an asymptotic state with a constant
 density in radius,  the frozen-in  magnetic field has a larger 
 magnetic flux densities in the
 outer region. In another word, the final
 magnetic field configuration has an inverted profile of
 the initial matter density distribution.

\begin{figure}
 \begin{center}
  \centerline{\includegraphics[width= 15 cm,height=19 cm]{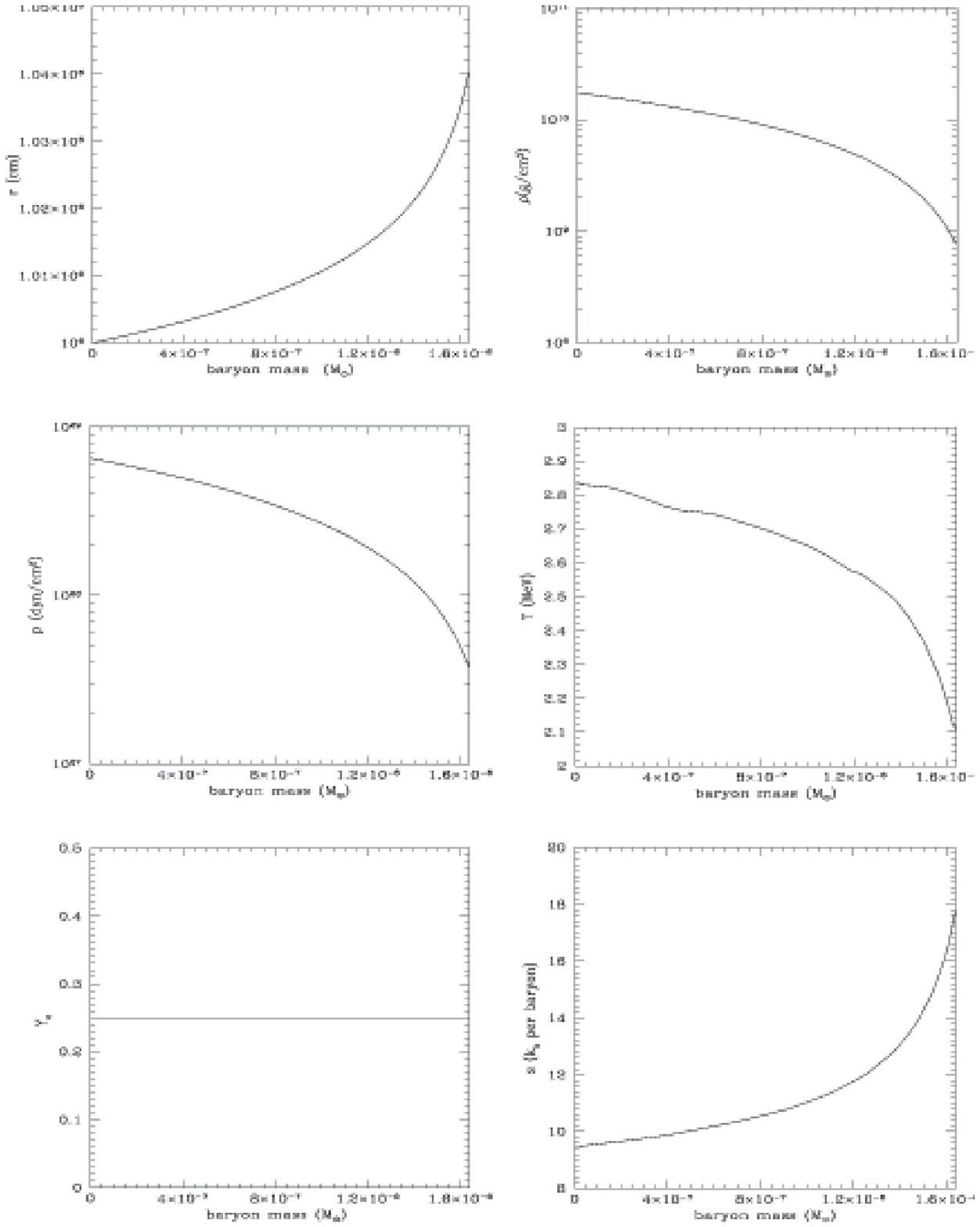}} 
 \end{center}
 \caption{The initial configuration of the surface layer.
 Radius $r$, baryon
 mass density $\rho$, matter pressure $p$, temperature $T$, electron
 fraction $Y_{e}$, and
 entropy per baryon $s$ are given as a function of baryon mass coordinate.}
 \label{fig:2}
\end{figure}

 This can be understood more precisely from equation (\ref{Bs}) and figure
 \ref{fig:2}. By using the same
 approximation that we have made previously, we
 can approximate
 equation (\ref{Bs})  as
\begin{equation}
B_{\theta}(t_1,m) \thickapprox
  \frac{r(t_1,m)}{r(t_0,m)}\frac{\rho(t_1,m)}{\rho(t_0,m)} B_{\theta}(t_0,m) ,
 \label{Bs2}
\end{equation}      
 for arbitrary $t_0$ and $t_1$. Taking $t_0$ and $t_1$, respectively, as the initial time and the time 
 at which the asymptotic state is reached, we find that
 $r(t_0,m)$, $r(t_1,m)$, and $\rho(t_1,m)$ depend on $m$ very weakly.
 $B_{\theta}(t_0,m)$ is also identical  on all mesh points. 
 Thus, the ratio of magnetic pressures between two 
 mesh points in the  asymptotic state can be approximated as
\begin{equation}
\frac{B_{\theta}(t_1,m_1)^2}{B_{\theta}(t_1,m_2)^2} \thickapprox
 \frac{\rho(t_0,m_2)^2}{\rho(t_0,m_1)^2}  .
   \label{Bs3}
\end{equation} 
 From the initial density distribution shown in figure \ref{fig:2} and the
 evolutions of density and magnetic pressure shown in figures
 \ref{10^6msev}-\ref{200-1msev}, this relation is verified.

 As a result of 
 the initial expansion and the later deceleration of the flow
 mentioned above,  
 the key quantities for the r-process except for the electron fraction
 show only a little change among the models. On the other hand, this
 change depends on the initial location of mass element
 in the atmosphere, which can be roughly divided into
 (i) the outer region, (ii) the  central region and 
(iii) the inner region as follows:
\newline
 (i) The outer region  ($1.2\times 10^{-6}M_{\odot} \lesssim m \lesssim 1.6\times 10^{-6}M_{\odot}$)
\newline
 Since the outer region has the lowest plasma beta, 
 the early rapid expansion occurs most 
 efficiently. Consequently, as matter passes
 through the heating region rapidly, the resulting entropy $s$  becomes lower 
 compared with the non-magnetic wind. This effect can be seen by
 comparing figure \ref{nonmsev} with figures \ref{10^6msev}-\ref{200-1msev}.
 Moreover, the deceleration takes place
 earlier on in the outer regions because the
 density gradient is initially
 larger than  the inner regions. 
 This occurs  during the $\alpha$-process stage, and makes the
 dynamical time scale $\tau_{dyn}$  longer. 
\newline
 (ii) The central region ($ 4.0\times 10^{-7}M_{\odot} \lesssim m
 \lesssim 1.2\times 10^{-6} M_{\odot}$)
\newline
 Though their plasma beta is not so low as in the outer region, the
 initial rapid
 expansion takes place also in this region and the entropy $s$ becomes
 lower than for the non-magnetic wind.
 Unlike the outer region, however, the deceleration occurs much later after the
 wind has already passed through the $\alpha$-process region ($\sim$50km
 to $\sim$150km). Thus, the  dynamical time scale becomes shorter.
 The three key quantities in the central region are displayed for all
 the models in  Table \ref{table:3}, in which
 we have taken   average  by excluding the outer
 and  inner 25 mesh points.
 The reduction of the dynamical time scale in the models with  strong magnetic
 fields  can be observed in the table.  However,
 the reduction is very small and, moreover, the
 entropy decreases at the same time. Thus,
 even if we  focus only on this 
 region, a successful r-process will  not take place. 
\begin{table}[h]
 \begin{center}
  \begin{tabular}{ccccc}
 \hline
 \hline
  Model  & $s(\textstyle{k_b})$ &  $\tau_{dyn}(\textstyle{s})$  & $Y_{e}$  \\[2pt]

 \hline
         &&&&\\[-10pt]

    N    &   130   & 3.37$\times$10$^{-2}$   &   0.453\\[3pt]

 10$^6$M & 130  &  3.37$\times$10$^{-2}$ & 0.453 \\[3pt]

 10$^3$M & 130  & 3.37$\times$10$^{-2}$ & 0.453 \\[3pt]

 10M  & 130   & 3.37$\times$10$^{-2}$  & 0.453 \\[3pt]

  1M  &  130  & 3.35$\times$10$^{-2}$  & 0.453 \\[3pt]

 $\frac{\textstyle{1}}{\textstyle{10}}$M  & 129    &  3.36$\times$10$^{-2}$   & 0.453 \\[3pt]

  $\frac{\textstyle{1}}{\textstyle{20}}$M  & 129    & 3.33$\times$10$^{-2}$  &  0.453 \\[3pt]

  $\frac{\textstyle{1}}{\textstyle{50}}$M  &  133    & 3.08$\times$10$^{-2}$  & 0.453 \\[3pt]

  $\frac{\textstyle{1}}{\textstyle{100}}$M  & 124     & 2.83$\times$10$^{-2}$  & 0.453 \\[3pt]

  $\frac{\textstyle{1}}{\textstyle{200}}$M  &  121   & 2.78$\times$10$^{-2}$  & 0.453 \\[3pt]

 \hline  
  \end{tabular}
 \end{center}
\caption{The three key quantities for each model. The quoted values are
 the average over the region excluding the inner  and 
 outer 25  mesh points. }
\label{table:3} 
\end{table}
\newline
 (iii) The inner region  ($m \lesssim 4.0\times 10^{-7} M_{\odot}$)
\newline
 The inner region  also experiences the acceleration  and
 deceleration stages. However, because of their large initial plasma
 beta, their effects are too small and  
 the key quantities for the r-process are essentially unaffected.

 As mentioned previously in \S2, our calculations approximate the flow
 near the equator with  initial magnetic fields perpendicular to the radial
 direction. Our results can be applied to the places in the wind
 where the deviation from the radial motion is small, that is, regions
 where
 the radial force  dominates over the transverse force. 
 The angular range that satisfies this condition can be
 estimated as follows.  For an arbitrary initial field-configuration,
 if the transverse velocity is
 small compared with the radial velocity, the magnetic fields are
 approximately determined by equations (\ref{Br})-(\ref{Bph}). 
 In the nonrelativistic limit, these equations become
\begin{equation}
B_{r}(t_1,m) = \frac{r^{2}(t_0,m)}{r^{2}(t_1,m)}B_{r}(t_0,m) ,
 \label{Br2}
\end{equation}     
\begin{equation}
B_{\theta}(t_1,m) = \frac{r(t_1,m)\rho(t_1,m)}{r(t_0,m)\rho(t_0,m)}B_{\theta}(t_0,m) ,
 \label{Bs2}
\end{equation}      
\begin{equation}
B_{\phi}(t_1,m) = \frac{r(t_1,m)\rho(t_1,m)}{r(t_0,m)\rho(t_0,m)}B_{\phi}(t_0,m) . 
 \label{Bph2}
\end{equation}      
 The evolution of the magnetic fields is determined by the initial
 magnetic fields,  radius, and density, as well as by the present
 radius and  density.
 The typical values  for the initial radius and
 density, and the final radius and density
 in  our models are
 10km, 10$^{10}$gcm$^{-3}$, 500km, and 2$\times$10$^3$gcm$^{-3}$,
 respectively.  
 If the initial  magnetic
 field is  uniform and perpendicular to the the equatorial
 plane,
 the radial force dominates over the transverse
 force roughly in the angular range between
 $65^{\circ} \lesssim \theta \lesssim 115^{\circ}$.
 This occupies a significant fraction of the
 total solid angle,
 and our results
 are applicable to a rather large region below and above the equatorial plane.

\section{Summary}
 In this paper we have investigated the effects of magnetic fields  on the
 neutrino-driven wind  and  explored the possibility as an r-process
 site  by performing  numerical
 simulations.
 The homogenous distribution is assumed for the initial magnetic fields,
 and  ten models with a
 wide range of field-strength have been computed.

 The results do not give the dynamical behavior expected  by
 Ref.~\cite{rf:9} \ 
 This may be the
 consequence of the assumed wind configuration, which   has a
 tendency to  increase the plasma beta rapidly and  
 the matter-pressure is dominated 
 in most of the wind evolutions. Hence, the magnetic pressure 
 has little  chance to influence  the dynamics.

 The evolutions of
 weak magnetic field models are identical with non-magnetic model. 
 The models with very strong magnetic fields
 ($B\gtrsim 10^{15}$G) show some, though not large,
 variation. For  about $0.5\times 10^{-6} M_{\odot}$ of wind, 
 the reduction of the dynamical time scale has been found. 
 However, this change is  very small and accompanied by the
 decreases of entropy. 
 Thus, unfortunately,
 the situation favored for a successful
 r-process is not realized in our models, even
 with magnetar-scale field-strengths.

 Since our study has been limited to initially
 homogeneous magnetic fields, other distributions of
 magnetic field should be investigated. Note, however, that the initial
 surface structure will be changed substantially then, since the
 magnetic pressure will determine the hydrostatic force balance.
 Multi-dimensional MHD simulations of magnetar formation
 may shed a light on this issue. 
 It is  also self-evident that more realistic and multidimensional
 investigations will be needed further, since the field-tension that is not
 considered in this paper may play an 
 important role. Even in that case, very strong magnetic fields
 $\gtrsim 10^{15}$G will be needed to influence the wind dynamics. 
 Last but not least, magnetic reconnections might be interesting, since
 the dissipation of magnetic fields will heat the matter, accelerating
 the wind. However, this is much beyond the scope of this paper.

\section*{Acknowledgements}

 Some of the numerical simulations were done on the supercomputer
 VPP700E/128 at RIKEN and VPP500/80 at KEK (KEK Supercomputer Projects
 No.108). This work was partially supported by the Grants-in-Aid for the
 Scientific Research (14740166, 14079202) from Ministry of Education,
 Science and Culture of Japan and by Grants-in-Aid for the 21th century
 COE program ``Holistic Research and Education Center for Physics of
 Self-organizing Systems''.

%

\begin{thebibliography}{99}
  
\bibitem{rf:1}
  S.~E.~Woosley, J.~R.~Wilson, G.~J.~Mathews, R.~D.~Hoffman and
	B.~S.~Meyer, \AJ{433,1994,229} 








\bibitem{rf:2}
  J.~Witti, H.~-Th.~Janka and Takahashi, A\&A \textbf{286} (1994), 841   

\bibitem{rf:3}
  K.~Takahashi, J.~Witti and H.~-Th.~Janka, A\&A \textbf{286} (1994), 857   

\bibitem{rf:4}
  Y.~-Z.~Qian and S.~E.~Woosley, \AJ{471,1996,331} 


\bibitem{rf:5}
  C.~Y.~Cardall and G.~M.~Fuller, \AJ{486,1997,111} 

\bibitem{rf:6}
  K.~Otsuki, H.~Tagoshi, T.~Kajino and S.~Wanajo, \AJ{533,2000,10}

\bibitem{rf:7}
  S.~Wanajo, T.~Kajino, G.~J.~Mathews and K.~Otsuki, \AJ{554,2001,578}
 
\bibitem{rf:8}
  K.~Sumiyoshi, H.~Suzuki, K.~Otsuki, M.~Terasawa and S.Yamada, PASJ
	\textbf{52} (2000), 601  

\bibitem{rf:8-5}
  S.~Nagataki and K.Kohri, PASJ
	\textbf{53} (2001), 547  


\bibitem{rf:9}
  T.~A.~Thompson, \AJ{585,2003,33}

\bibitem{rf:10}
  S.~Yamada, \AJ{475,1997,720}

\bibitem{rf:11}
  C.~W.~Misner and D.~H~.Sharp, \PRB{136,1964,571}

\bibitem{rf:12}
  A.~Achterberg, \PRA{28,1983,2449}

\bibitem{rf:13}
  H.~Shen, H.~Toki, K.~Oyamatsu and K.~Sumiyoshi, \NPA{637,1998,435}

\bibitem{rf:14}
  H.~Shen, H.~Toki, K.~Oyamatsu and K.~Sumiyoshi, \PTP{100,1998,1013}


\bibitem{rf:15}
  Y.~-Z.~Qian, Prog. Part. Nucl. Phys. \textbf{50} (2003),153  


\bibitem{rf:16}
  R.~D.~Hoffman, S.~E.~Woosley and Y.~-Z.~Qian,  \AJ{482,1997,951} 

\bibitem{rf:17}
  M.~Terasawa, K.~Sumiyoshi, S.~Yamada, H.~Suzuki and T.~Kajino,
  \AJ{578,2002,137} 

\bibitem{rf:18}
  C.~Sneden, A.~McWilliam, G.~W.~Preston and J.~J.~Cowan,
  \AJ{467,1996,819} 

\bibitem{rf:19}
  Y.~Ishimaru and S.~Wanajo,
  \AJ{511,1999,33} 

\bibitem{rf:20}
  T.~Tsujimoto, T.~Shigeyama and Y.~Yoshii,
  \AJ{519,1999,63} 

\bibitem{rf:21}
  S.~Wanajo, N.~Itoh, Y.~Ishimaru, S.~Nozawa and T.~C.~Beers,
  \AJ{577,2002,853} 

\bibitem{rf:22}
  D.~Argast, M.~Samland, F.~-K.~Thielemann and Y.~-Z.~Qian, 
   A\&A \textbf{416} (2004), 997
\end{thebibliography}
\end{document}